\documentclass[sn-aps]{sn-jnl}

\usepackage{graphicx}
\usepackage{xcolor}
\usepackage{caption}
\usepackage{subcaption}
\usepackage{multirow}
\usepackage{amsmath,amssymb}
\usepackage[numbers,sort&compress]{natbib}
\usepackage{threeparttable}
\usepackage{textcomp}
\title[Imaging by a single photon sensitive camera]{Imaging of CsI(Tl) crystal event and double-slit Young's interference by a single photon sensitive camera}
\author*[1,2,3]{Zhimin Wang}\email{wangzhm@ihep.ac.cn}
\author[1,2]{Min Li,}
\author[1,2]{Diru Wu,}
\author[1,3]{Jinchang Liu,}
\author[1]{Xiangcheng Meng,}
\author[1,2]{Caimei Liu,}
\author[1,2]{Changgen Yang}
\affil[1]{Institute of High Energy Physics, Beijing 100049, China}
\affil[2]{University of Chinese Academy of Sciences, Beijing 100049, China}
\affil[3]{State Key Laboratory of Particle Detection and Electronics}
\abstract{We will discuss an imaging measurement with a single photon sensitive and low noise camera aiming to a new paradigm in the optical readout of scintillation detectors. The features of the single photon sensitive camera will be characterized and demonstrated with a measurement on double-slit Young's interference in single photon mode. An imaging test on CsI(Tl) crystal and alpha source will be performed further for preliminary measurements on the noise level and sensitivity of the system with a 1/2", f/1.4 lens, which reaches an sensitivity on light intensity around 1/10 of the 3-inch PMT and shows a potential to realize an imaging of single alpha event. An application proposal to scintillation detectors will be further discussed, where it is usually assumed that the imaging is not possible in such a photon-starved and large-emittance regime.
}


\date{Received: date / Accepted: date}

\begin{document}
\maketitle
\flushbottom

%\appendix
%\section{Some title}
%Please always give a title also for appendices.

\section{Introduction}
\label{sec:1:intro}

Photon detection is one of the fundamental technologies widely used in particle physics, astronomy and industry, such as Super-K\cite{superK-FUKUDA2003418}, KamLAND\cite{KamLAND-PhysRevLett.90.021802}, JUNO\cite{JUNOdetector,Abusleme2022MassTA-JUNO-PMT}, Hubble\cite{Hubble-10.1117/12.324464}, clinically and preclinically\cite{Single-Photon-Zanzonico2016}, X-ray imaging\cite{Single-photon-CCD-2014} and long-range imaging\cite{3Dimaging-45km-Li:20} etc.
The reconstruction of vertex and track is critical in most of the particle physics experiments, where it is always trying to use most of information from photons.
Ignoring polarization, each photon can be described by six coordinates when it is received: its impact position on the photodetector surface (x,y), the time of arrival t, two directions of propagation, $\theta$ and $\phi$, and the wavelength $\lambda$\cite{LS-imaging-PhysRevD.97.052006}. Photodetectors, typically PMT or SiPM in particle physics experiments, are commonly used for light strength and timing measurements and some crude event localization. Generally, a full coverage of photodetector assemblies to the whole target guarantees an overall light collection efficiency for better precision measurement as proposed in \cite{JUNOdetector}.

The search for a novel technology able to detect and reconstruct events in scintillation detectors has become more and more important for dark matter and neutrino studies. An idea of detecting the light, optical and directional readout approach, proposed many years ago \cite{DOMINIK1989779}, has received renewed attention in recent years, where the challenges are the needs of high spatial resolution over large volumes, limited signal strength and non-negligible noise level\cite{BATTAT20161}. The charge coupled devices (CCD) have been widely used in the past as high granularity light sensors and an upgrade of classical emulsion radiograph, such as GEM-based TPC with a 2-D CCD readout for directional dark matter experiments\cite{PHAN201682,Deisting2021CommissioningOA,TPC-CCD-Phan_2020,TPC-CCD-FRAGA2002357,TPC-CCD-2014}, thermal neutron imaging\cite{MOR2021165632}, single photon counting X-ray CCD camera spectrometer\cite{Single-photon-CCD-2014}, photonic graph states\cite{Multiphoton-Graph2020}, transparency measurement\cite{XIE2021165459}, a kilogram-scale Skipper CCD to detect coherent elastic neutrino nucleus scattering\cite{skiper-CCD-2021,skipper-snowmass,PhysRevD.91.072001}, and classical emulsion radiograph replaced by digital detector imaging especially in medicine applications\cite{Schumacher2016PhotonCA}.
At the same time, some other similar devices are also developed and used for single-photon light detection and ranging (lidar)\cite{3Dimaging-45km-Li:20,SPC-v2}, and single image 3-D photography\cite{jampani2021slide}.

But, a critical limitation of CCD is its high level of readout noise up to one to tens electrons per pixel (RMS) comparing to the signal strength in photon counting level of particle physics experiments.
Bubble chamber as an example, its imaging can be realized via appropriate wide angle lenses with finite aperture, but external illumination is essential to provide sufficient track brightness over noise to produce a conventional image (efficient signal-to-noise ratio). Furthermore, it is commonly concluded that the scintillation detectors are fundamentally unsuitable for the imaging applications owing to the photon-starved regime and uniform angular distribution of the light produced in the scintillation process, where the limited photons emitted over a large angular range cannot be imaged through a modest aperture optical system under the limitation of the sensor noise.
Homogeneous scintillation detectors give up imaging entirely, as the price to pay to collect a large fraction of the emitted photons by maximizing the coverage with low noise photon sensors, and some equal options are discussed such as "distributed imaging" with PMTs in \cite{LS-imaging-PhysRevD.97.052006}.

More recently, cameras based on active pixel sensor (APS) technology developed on complementary metal-oxide semiconductors (CMOS) have been developed that can reach tens of millions of pixels with sub-electron readout noise and single photon sensitivity (usually referred as scientific CMOS (sCMOS)), where imaging, single photon imaging in particular, is providing another new possibility. It could overcome the main limitation of the CCDs on noise used for scintillation detector to offer several advantages: highly performing optical sensors can be easily procured and being developed for commercial applications; the use of suitable lenses allows the possibility of imaging large area with few sensors and better spatial resolution.
A combined system of the low-noise and high-granularity cameras and fast light sensors will provide more powerful technologies and possibilities for better reconstruction of vertex and track, and benefit for a better particle identification\cite{CYGNO-instruments6010006-2022}.

In this paper, we will show a preliminary measurement on a single photon sensitive and low noise camera with CsI(Tl) crystal. Sec.\ref{sec:1:setup} will introduce the setup of the camera testing system. Sec.\ref{sec:1:photography:sp} will discuss the characterization of the camera following a measurement with single photon imaging of double-slit Young's interference. Sec.\ref{sec:1:crystal} will have a preliminary test on CsI(Tl) crystal and alpha source with the camera system. Sec.\ref{sec:1:dis} will provide a proposal on spatial coincidence measurement by cameras associating with real time photon sensors, and a short summary in Sec.\ref{sec:1:summary}.

\section{Imaging setup}
\label{sec:1:setup}

The single photon sensitive and low noise camera of ORCA-Quest qCMOS C15550-20UP is a new product of Hamamatsu Photonics\,\cite{HPK-ORCA-QUEST} with a rolling shutter, 4096\,(H) $\times$ 2304\,(V) pixels (4.6$\times$4.6\,\textmu m$^2$/pixel), a low-dark current to 0.006 electron/pixel/s and an ultra-low readout noise to 0.27 electron rms under ultra quiet scan. The camera can be coupled with lenses through C type interface. It will work under air cooling mode to keep the chips temperature at -20\,$^{\circ}$C here.
For our study, a combined system except the camera is designed with 3-inch PMTs\,\cite{JUNO-3inch-CAO2021165347} which are working at a gain of $3\times10^6$ with a $\times$10 fast amplifier to cover single photon level, and acquired by a CAEN digitizer DT5751\,\cite{CAEN-dt5751} for signal intensity monitoring in particular. The system will be re-configured manually in the following measurements including pulsed LED, crystal and double-slit Young's interference.

%, and here a lens of f=25\,mm, F1.4, C 2/3", 2MP is used

\begin{figure*}[!hbt]
    \centering
    \includegraphics[scale=0.25]{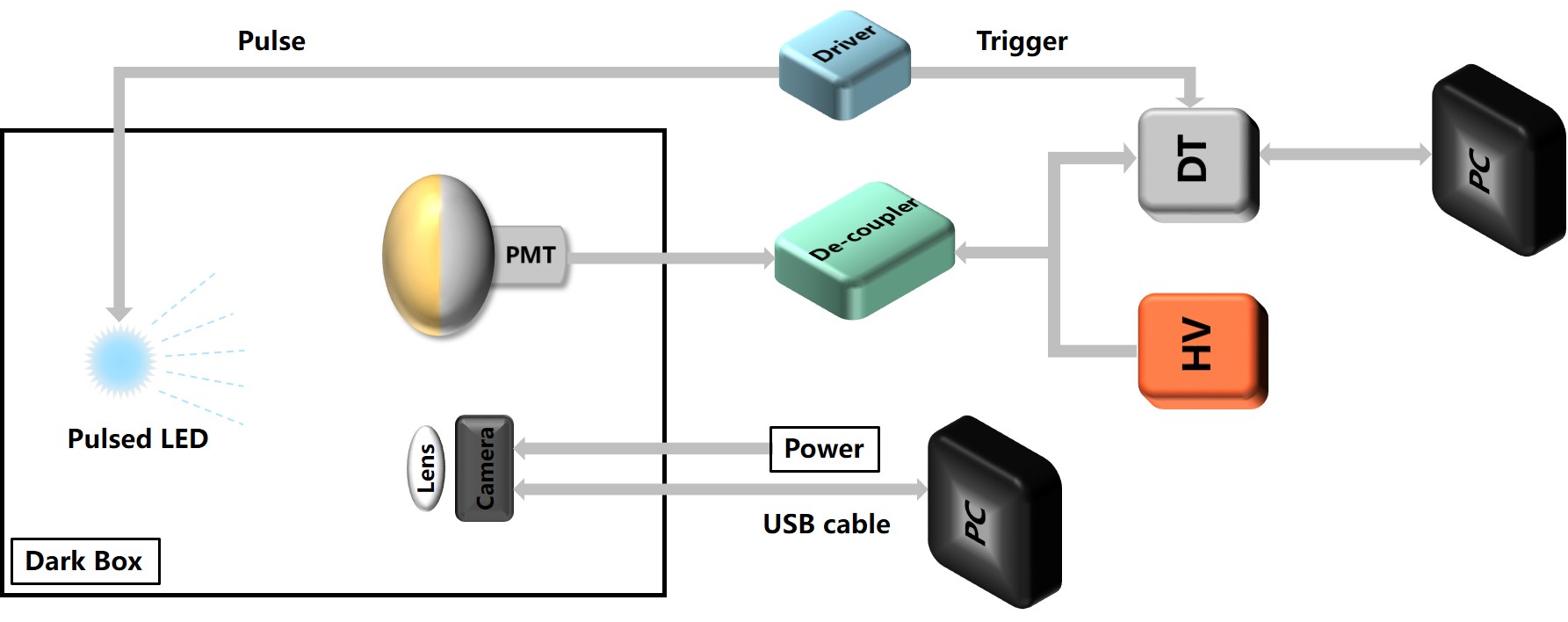}
    \caption{Schema of LED testing system.}
    \label{fig:darkbox-schema}
\end{figure*}

Fig.\ref{fig:darkbox-schema} shows the setup for pulsed LED measurement.
The PMT (also the camera) is located at a distance ($\sim$15\,cm) to the light source to monitor the light intensity of the LED pulses. The LED is encased in a nylon diffuser ball to illuminate uniformly the PMT and the camera. The LED is powered by a driver pulse with 30\,ns width under different frequency, and its wavelength is around 420\,nm. The light intensity viewed by the PMT (and the camera) can be adjusted from single photon to multi-photons through tuning the driver amplitude.
The camera can be configured manually to different exposure time. The data of the camera is saved in tagged image file format (tif) with 16\,bits per pixel, and each image is 18\,MB normally.

\begin{figure}[!htb]
	\centering
	\begin{subfigure}[c]{0.45\textwidth}
	\centering
	\includegraphics[width=\linewidth]{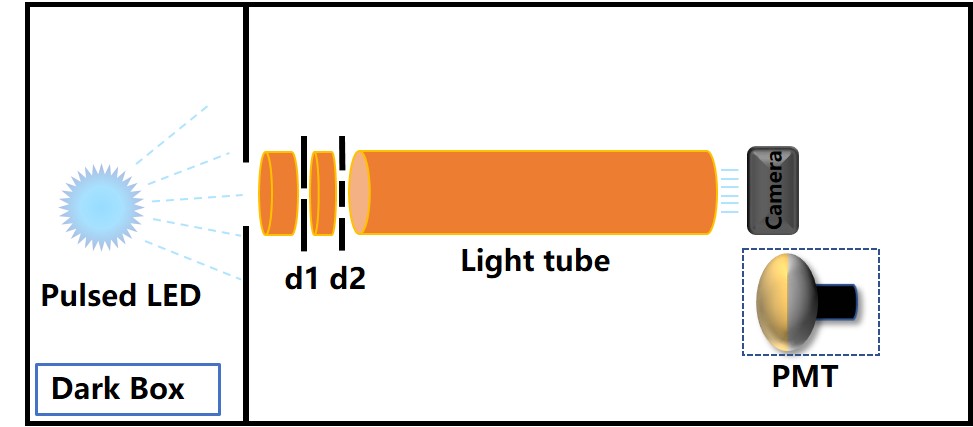}
    \caption{Schema of Young's interference.}
	\label{fig:darkbox-yang}       % Give a unique label
	\end{subfigure}	
	\begin{subfigure}[c]{0.338\textwidth}
	\centering
	\includegraphics[width=\linewidth]{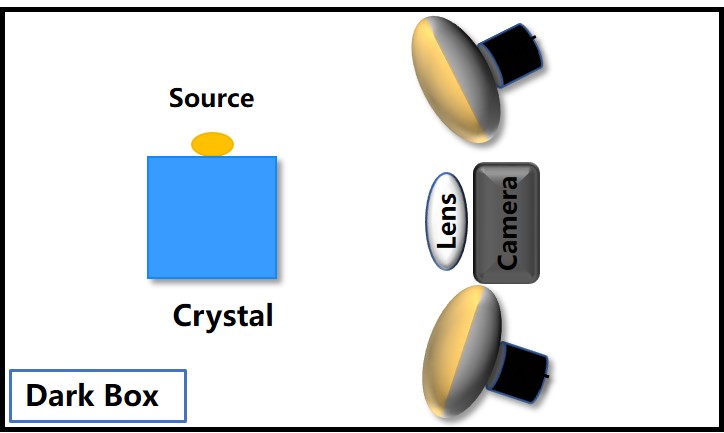}
    \caption{Schema of crystal test.}
	\label{fig:darkbox-crystal}       % Give a unique label
	\end{subfigure}	
    \caption{Re-configured systems for crystal and single photon double-slit Young's interference measurements.}
	\label{fig:darkbox-schema2}       % Give a unique label
\end{figure}

Thanks to the excellent ability of the system on light intensity control and imaging, it is exciting to realize a double-slit Young's interference measurement by single photon with the re-configured system as shown in Fig.\ref{fig:darkbox-yang}. The width of slit d1 and each slit of d2 is 0.1\,mm and 0.029-0.040\,mm respectively, and the distance between the slits of d2 is 0.2\,mm. The distance between slit d1 and d2 is 15\,cm, and between the slit d2 and the output end of the interferometer tube is 60\,cm, respectively. Another bare LED with wavelength around 500\,nm will be used to provide stronger light intensity in pulsed mode for the interference. The light intensity at the output end of the interferometer tube will be measured by the PMT, and in turn with the camera. The received light intensity on average by the PMT is set to around 0.1\,photon of each pulse following Poison distribution. The camera will be run without lens to view the interference fringe directly.

For a crystal measurement, the system is re-configured as shown in Fig.\ref{fig:darkbox-crystal} aiming for the imaging test on event vertex or track in scintillation detector. Here a CsI(Tl) crystal (light yield $\sim$45,000 photons/MeV\cite{CsI-1993-8526779,BELOGUROV2000254}) and $^{241}$Am alpha source (alpha energy 5.5\,MeV\cite{am241-alpha-BARANOV1963547}) are used to replace the LED and located at around 10\,cm in front of the PMTs and camera. The PMTs are used to monitor the signal intensity, the coincidence of which will be used to trigger the PMT waveform data taking.

\section{Single photon sensitive camera}
\label{sec:1:photography:sp}

\subsection{Single photon response and noise}
\label{sec:2:noise}

\begin{figure}[!htb]
	\centering
	\begin{subfigure}[c]{0.65\textwidth}
	\centering
	\includegraphics[width=\linewidth]{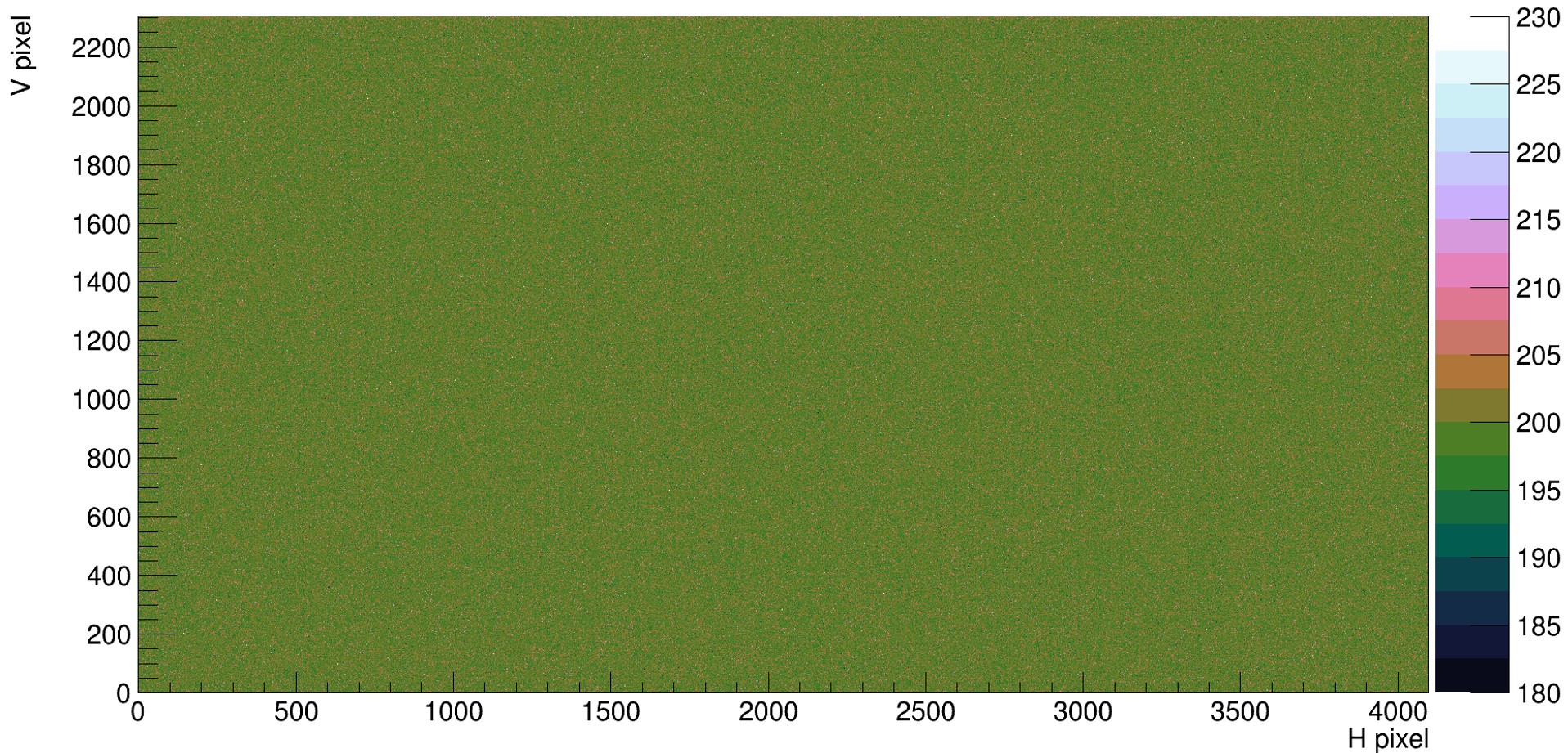}
	\caption{2-D image of the camera in dark w/ exposure 1\,s}
	\label{fig:noise:2D}       % Give a unique label
	\end{subfigure}	
	\begin{subfigure}[c]{0.45\textwidth}
	\centering
	\includegraphics[width=\linewidth]{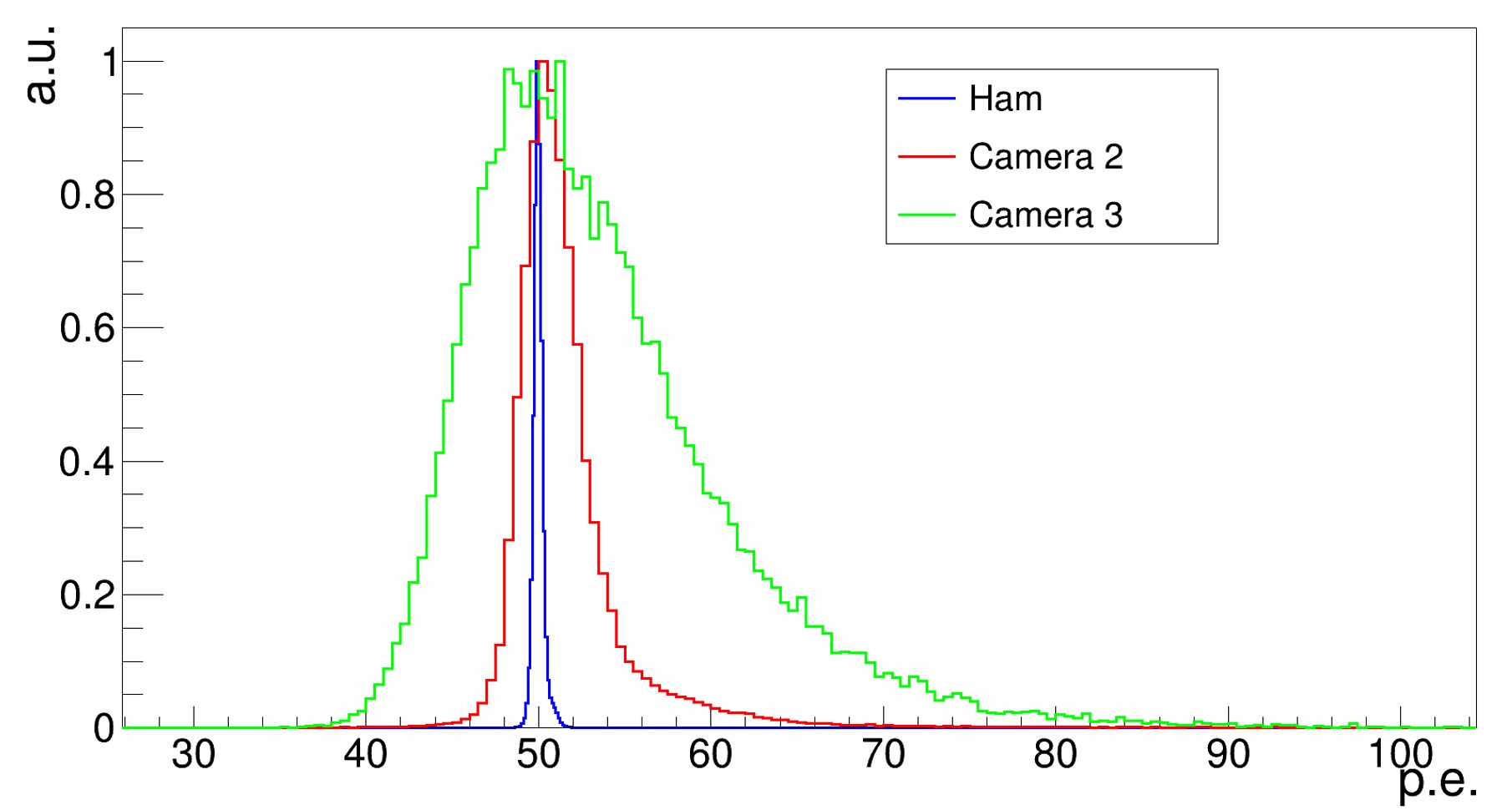}
    \caption{1-D plot of pixel noise in (a)}
	\label{fig:noise:1D}       % Give a unique label
	\end{subfigure}
	\begin{subfigure}[c]{0.45\textwidth}
	\centering
	\includegraphics[width=\linewidth]{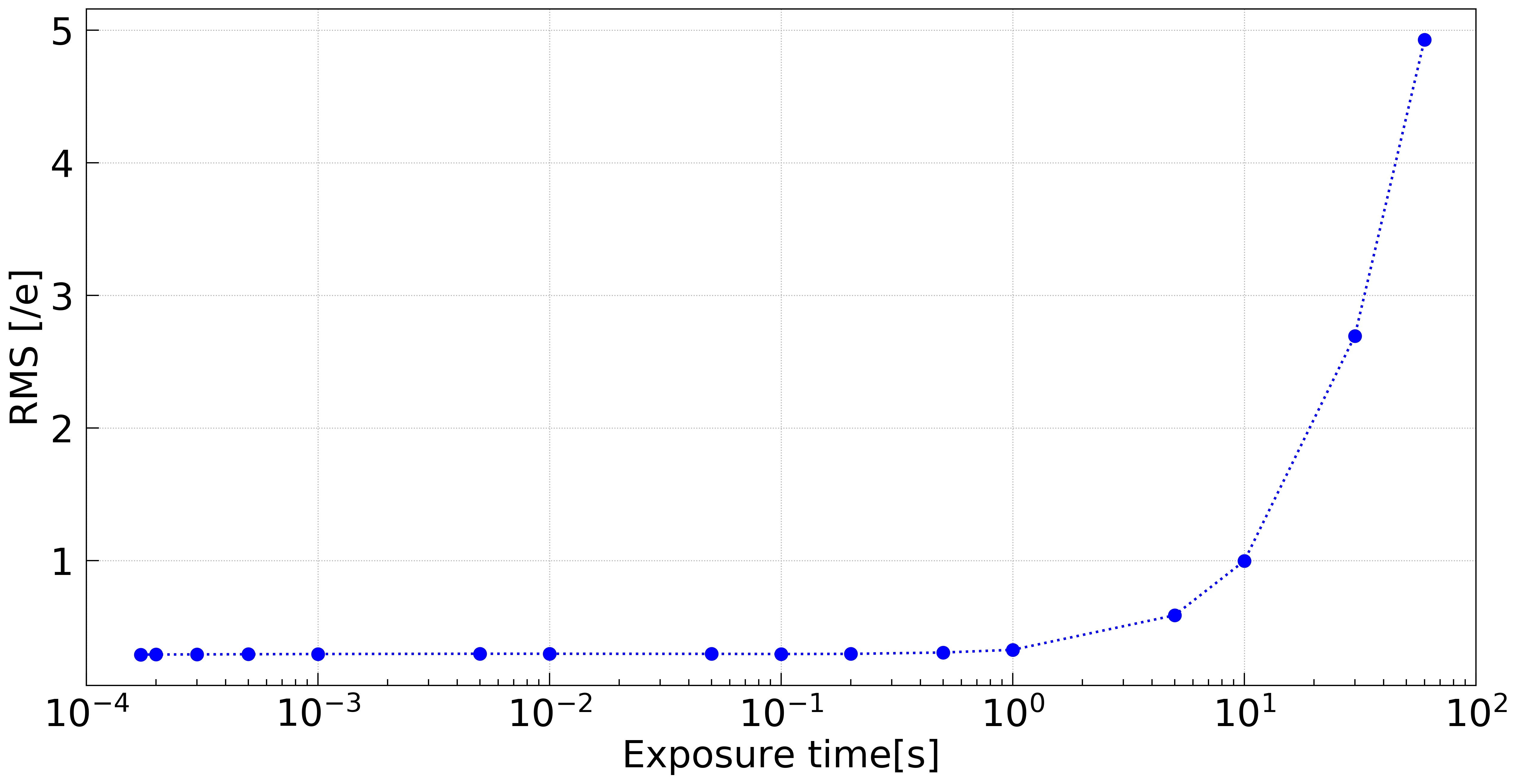}
    \caption{Noise vs. exposure time}
	\label{fig:noise:exposure}       % Give a unique label
	\end{subfigure}	
    \caption{Camera noise. (a) 2-D image of the camera with 1\,s exposure time (in dark), where the value of each pixel is in ADC (baseline is at 200\,ADC); (b) the 1-D distribution of pixel noise in p.e. of the Hamamatsu camera (blue), and other two kinds of cameras (red and green); (c) pixel noise of the camera vs. exposure time.}
	\label{fig:noise}       % Give a unique label
\end{figure}

The noise of the camera is measured in dark firstly, where the camera is configured in ultra-low noise mode (low frequency mode) as mentioned. Fig.\,\ref{fig:noise:2D} shows an example of the 2-D images taken by the camera with 1\,s exposure time, 9,437,184 pixels in total. Fig.\,\ref{fig:noise:1D} shows the 1-D noise plot of pixels in unit of photoelectron (p.e.)\footnote{Here, we will equally use the electron and photoelectron to evaluate the noise level or light intensity.}, where the baseline is artificially shifted to 50\,p.e., converted from the raw ADC by a factor of 7.8\,ADC/p.e. (the baseline in ADC is at 200). The standard deviation (rms) of the distribution of all the pixels will be used to evaluate the noise level of the camera. The noise of the Hamamatsu camera is much smaller than the other two typical cameras with the same exposure time.

The calculated noise level versus exposure time from 172\,\textmu s to 60\,s is shown in Fig.\,\ref{fig:noise:exposure}, where a linear fitting is applied. It is confirmed that the readout noise of pixels is around 0.28\,electron (around 2.2\,ADC), and the dark current noise, proportional to the exposure time, is around 0.08\,electron/pixel/s (around 0.62\,ADC), which mean an excellent performance even a little bit worse than the statement.
Considering the total 4096$\times$2304 pixels with 1\,s exposure time (the noise level around 0.36\,p.e./pixel), the pixel number over 0.5\,p.e. is around 478,203, which can be treated as dark count rate 478\,kHz as one channel, and much higher than the traditional PMT only in hundreds to thousands count per second. At the same time, there are also some pixels showing higher noise over 3.5\,p.e., which reaches to 713\,Hz.

\begin{figure}[!htb]
	\centering
	\begin{subfigure}[c]{0.4\textwidth}
	\centering
	\includegraphics[width=\linewidth]{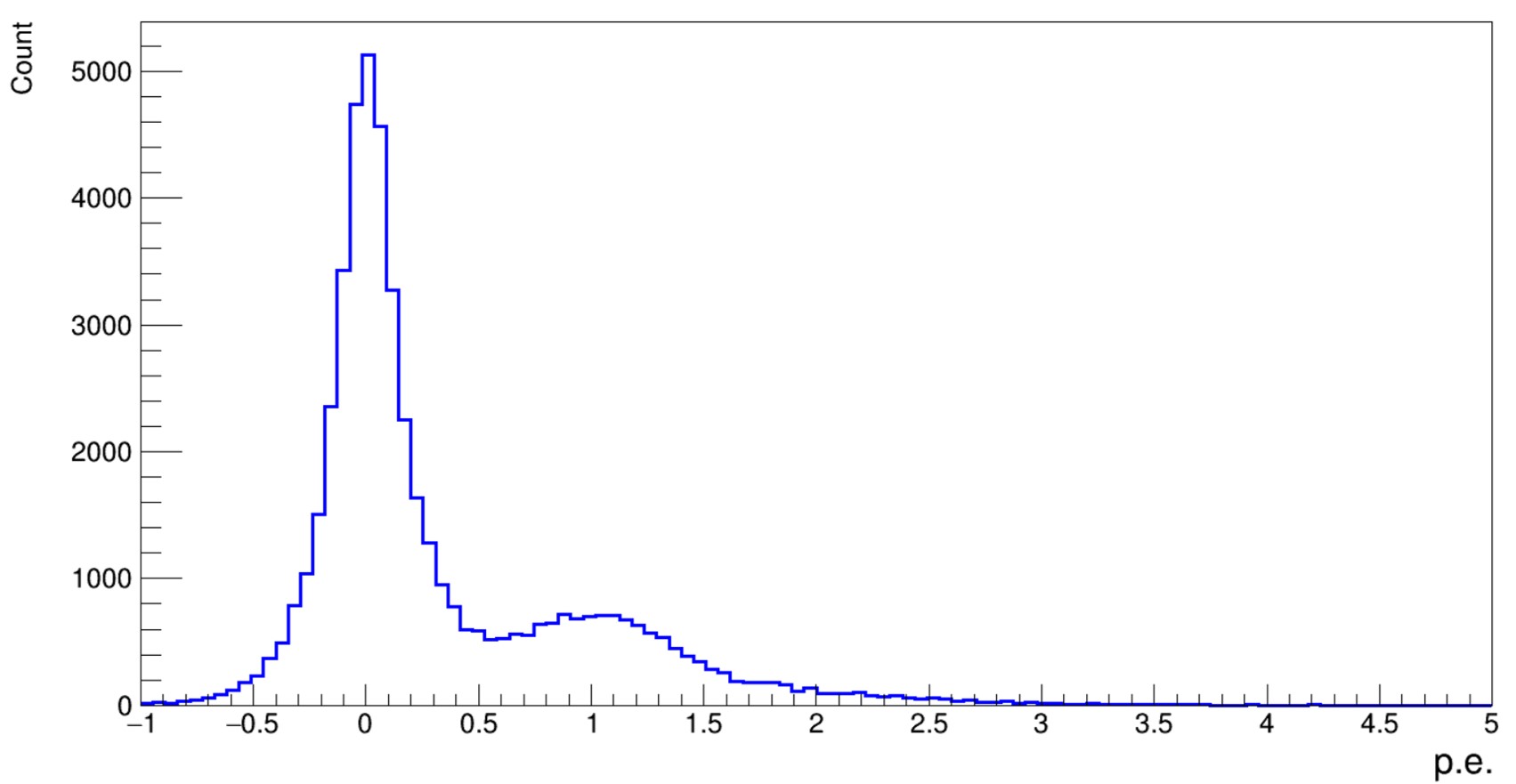}
    \caption{Intensity viewed by PMT.}
	\label{fig:camera:LED:PMT}       % Give a unique label
	\end{subfigure}
	\begin{subfigure}[c]{0.44\textwidth}
	\centering
	\includegraphics[width=\linewidth]{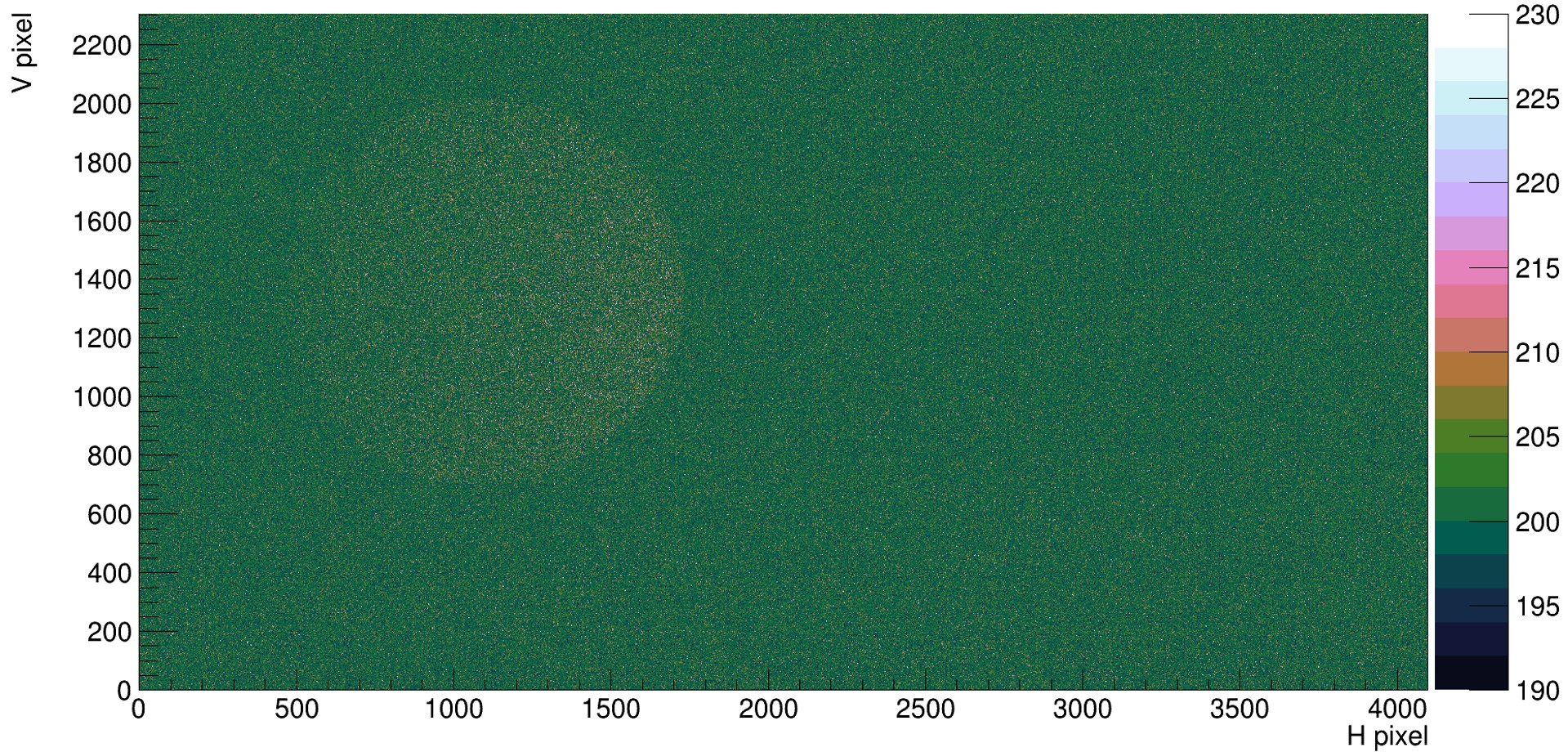}
    \caption{2-D image of LED}
	\label{fig:camera:LED:2D}       % Give a unique label
	\end{subfigure}
	\begin{subfigure}[c]{0.4\textwidth}
	\centering
	\includegraphics[width=\linewidth]{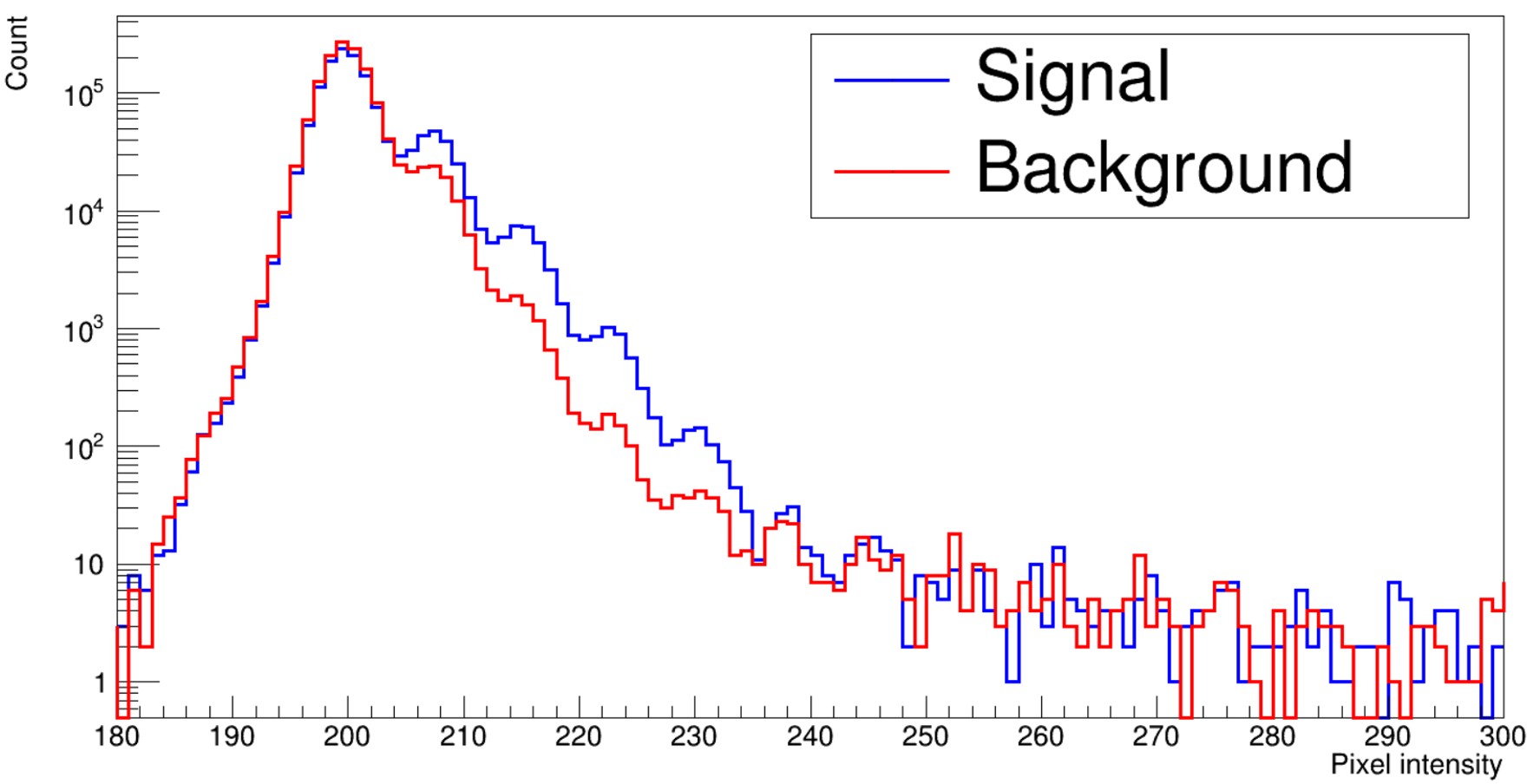}
    \caption{1-D plot of pixels intensity}
	\label{fig:camera:LED:1D}       % Give a unique label
	\end{subfigure}	
	\begin{subfigure}[c]{0.4\textwidth}
	\centering
	\includegraphics[width=\linewidth]{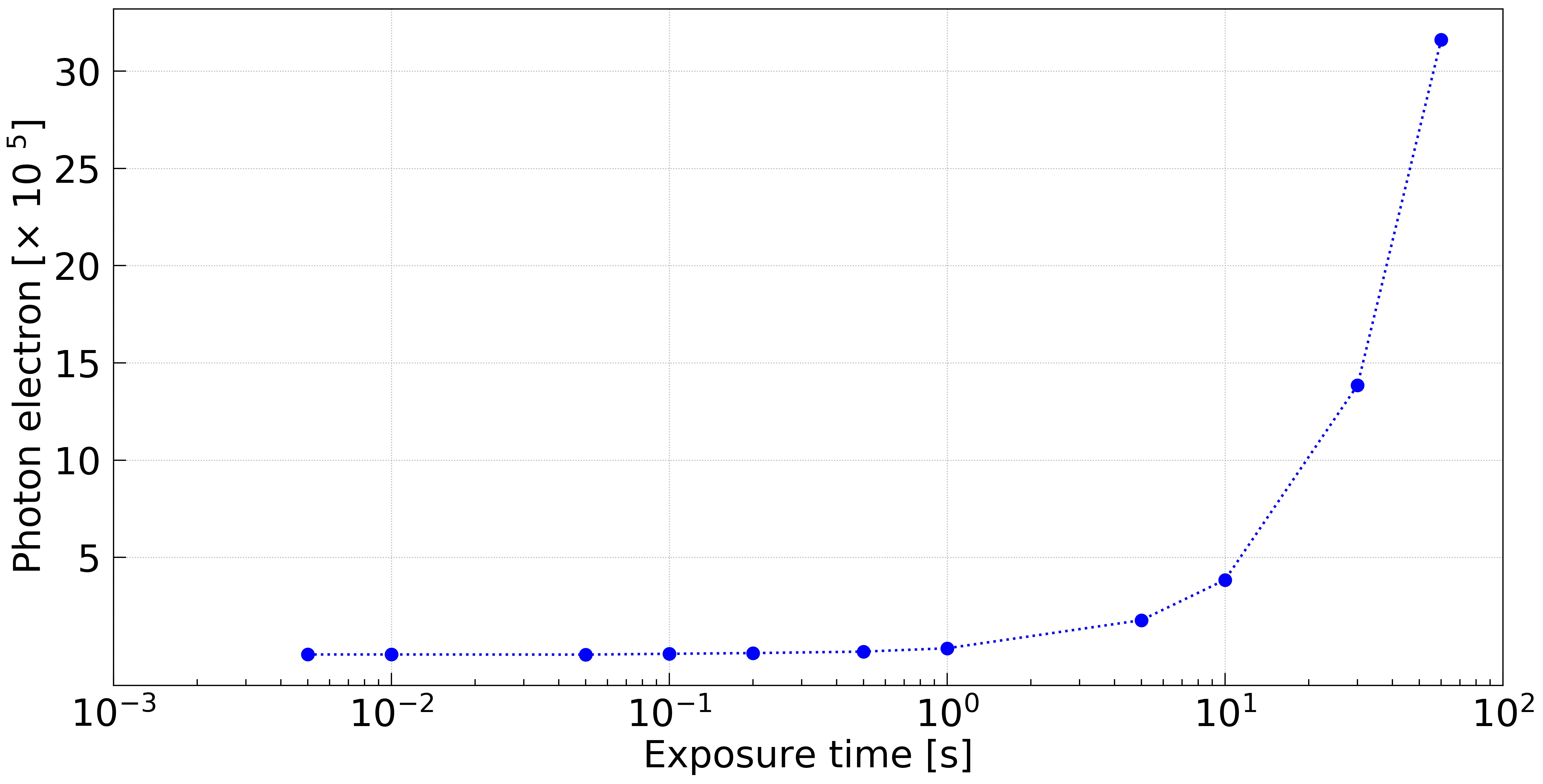}
    \caption{Intensity vs. exposure time.}
	\label{fig:camera:LED:camera}       % Give a unique label
	\end{subfigure}	
    \caption{Image testing with pulsed LED. (a) the intensity of LED pulse in p.e. monitored by PMT; (b) 2-D image of the LED with 5\,s exposure time, where the shape of diffuser ball can be identified obviously; (c) the 1-D intensity in ADC of all the pixels of the diffuser ball target region (blue) in (b), where the background (red) is from a shifted area; (d) the measured light intensity by the camera versus different exposure time.}
	\label{fig:camera:LED}       % Give a unique label
\end{figure}

A LED with diffuser ball diameter of 4\,cm is used to illuminate uniformly the 3" PMT and the camera coupled with a lens (2/3", C, 25-$\infty$\,mm, f/1.4, 2\,MP) as shown in Fig.\,\ref{fig:darkbox-schema}. The LED is driven by a positive rectangular pulse in 30\,ns duration. The LED intensity per pulse, in single photon level, is adjusted to around 0.34\,p.e. on average viewed by the 3" PMT according to Poisson distribution as shown in Fig.\,\ref{fig:camera:LED:PMT}, or around 1.5\,photons considering the PMT QE ($\sim$23\%).
Considering the quantum efficiency (QE) ($\sim$23\% and $\sim$85\%) at 420\,nm, the effective aperture (3" and 2/3", f/1.4), and the distance to LED ($\sim$18\,cm and $\sim$15\,cm) of the PMT and camera respectively, the expected light intensity per LED pulse viewed by the camera is around 1/9 (or 0.15\,p.e./pulse) to expected photons viewed by the PMT.
Fig.\,\ref{fig:camera:LED:2D} shows an image of the LED with 500\,kHz driver frequency taken by the camera with 5\,s exposure time. The center of the LED is located at pixel (1080, 1354) with a radius of around 675 pixels (the whole region contains about 1.4 million pixels). The measured dimension (3.7$\pm$0.3\,cm) is consistent with the diffuser ball considering the focal length and object spacing, where the magnification is around 1/6.

The 1-D intensity of all the pixels of the LED region is shown in Fig.\,\ref{fig:camera:LED:1D}, where the background is gotten from an equal area as the LED but with a shifted center at pixel (3080,1354). The measured strength of each pixel covers from single p.e. (around 208\,ADC with baseline at 200\,ADC) to several p.e.s, where individual p.e.s can be identified clearly. The LED is further imaged with different exposure time, and the measured light intensity after camera noise correction (based on the background region) is shown in Fig.\,\ref{fig:camera:LED:camera}. It is concluded that the light intensity viewed by the camera is around 52032\,p.e./s by a linear fitting, which derives around 0.10\,p.e./LED pulse or 0.04\,p.e./pixel with 1\,s exposure time on average. It is a little bit smaller than the expected value, but basically consistent considering the uncertainty of acceptance solid angle, QE and the transparency of the lens, which still needs careful validation for a precise quantitative relationship. But it is clear that the measured intensity by the camera is proportional to the exposure time exactly until much short exposure time. The minimum exposure time to identify the LED here is around 0.05\,s (integrated LED intensity to $\sim$2600\,p.e. or 25\,k pulses), when it is dominated by the camera noise of the LED region. Following the understanding, it is possible to identify a light source in single pulse from few to tens p.e.s by a smaller target region (for example a radius in 1\,mm or 1/400 of the current LED area), where the camera noise can be controlled to a limited level. It will be further discussed in the following sections.

\subsection{Double-slit Young's interference}
\label{sec:2:yang}

As known, the wave-particle duality is one of the fundamental features of quantum mechenics, which has been measured by a lot of experiments with photons, electrons and atoms, even in single particle level\cite{1954-doi:10.1063/1.1770945,single-electron-Matteucci_2013,Spence2009,PhysRev.90.490,review-PIPKIN1979281,Grangier-1986,single-electron-1989,interference-Scully1991,singlephoton-interference-RevModPhys.71.S274,electron-Bach-2013,electron-Sanz2014QuantumIA,Trajectories-Single-Photon-scince-2011}. The double-slit Young's interference measurement is one of the famous experiments in the physics history and excellent interpretation on quantum mechanics  \cite{double-slit-review-2002,double-slit-PhysRevA.95.042129,Feynman-1963,Young-Feynman-2019}.
While most of the previous measurements on double-slit Young's interference with single photon are configured with special light source, limited on the spatial resolution, single photon sensitivity and dynamic range of the photon sensors\cite{double-slit-review-2002,double-slit-PhysRevA.95.042129,Trajectories-Single-Photon-scince-2011,three-slit-PhysRevA.85.012101,three-slit-2010,three-slit-2016}. The newly developed single-photon sensitive camera, monitored by PMT in particular, provides an excellent opportunity to realize the double-slit interference fringe in single photon level to verify the basic quantum theory directly.
%Now, quantum entanglement is widely regarded as one of the most prominent features of quantum mechanics and quantum information science \cite{quatum-simulation-Aspuru-Guzik2012,photon-pair-2021}.

With a traditional Young's double-slit interferometer as shown in Fig.\,\ref{fig:darkbox-yang}, we will try to measure the interference fringe with single photon and normal LED source. The light source used here is another bare LED (3\,mm, $\lambda\sim$500\,nm) without diffuser ball and working in pulse mode by a pulse driver of 100\,ns width. The light intensity at the output end of the interferometer is adjusted to single photon level, which is around 0.02\,p.e./pulse on average viewed by the PMT and much smaller than that as shown in Fig.\,\ref{fig:camera:LED:PMT}. It means around 0.1\,photon/pulse on average considering the QE of the PMT, where the possibility of single photon per pulse is around 0.1, and more than 2\,photons per pulse is around 0.005, respectively. The relative stability of the light source monitored by PMT is around 5\%. The intensities viewed by PMT with or without the slits are further measured at the output end of the interferometer: around 7.7\,p.e./pulse without either slits d1 or d2, around 3.6\,p.e./pulse only with slit d1 under the same LED configuration.

\begin{figure}[htb]
	\centering
	\begin{subfigure}[c]{0.23\textwidth}
	\centering
	\includegraphics[width=\linewidth]{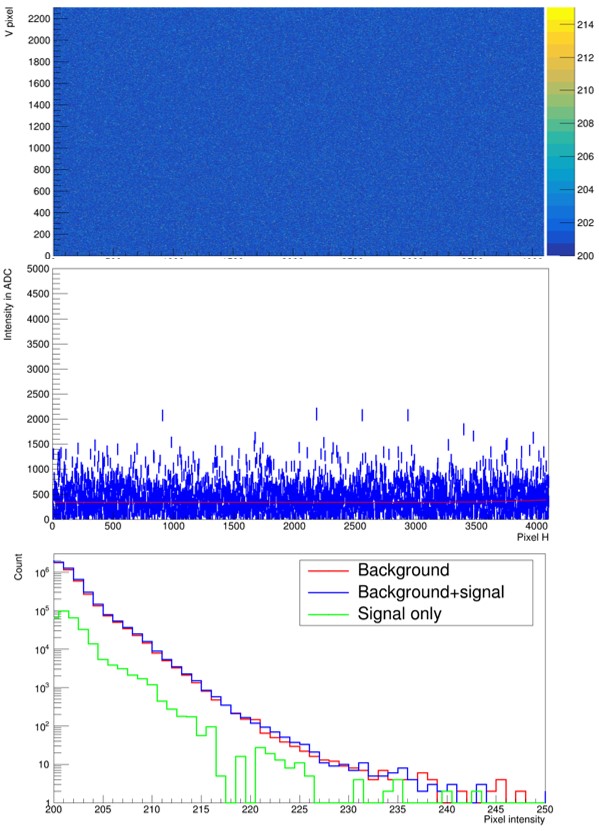}
    \caption{0.1\,s}
	\label{fig:LED-interference:1}       % Give a unique label
	\end{subfigure}	
	\begin{subfigure}[c]{0.23\textwidth}
	\centering
	\includegraphics[width=\linewidth]{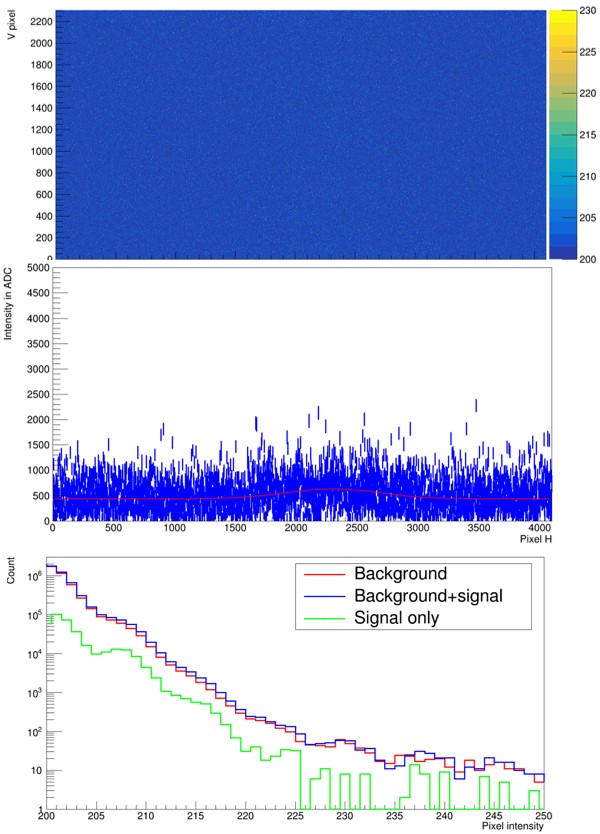}
    \caption{1\,s}
	\label{fig:LED-interference:2}       % Give a unique label
	\end{subfigure}	
	\begin{subfigure}[c]{0.23\textwidth}
	\centering
	\includegraphics[width=\linewidth]{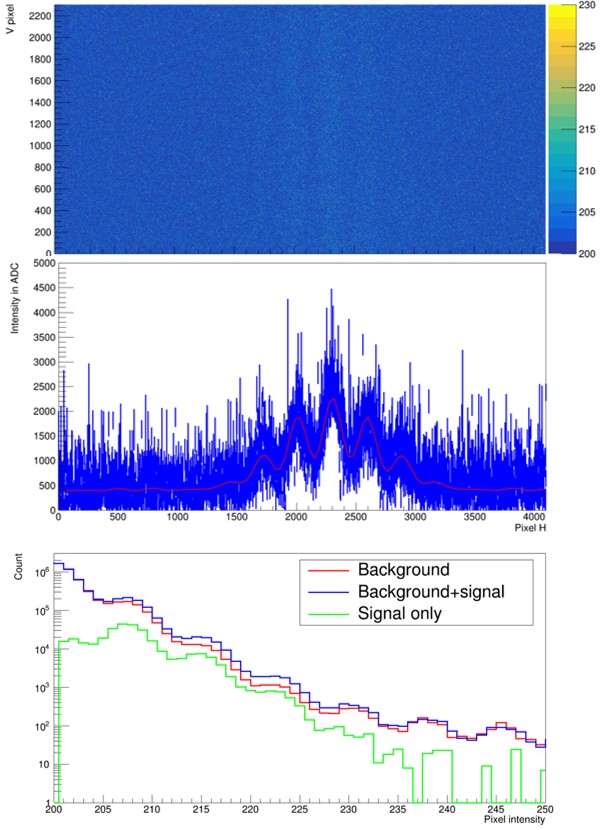}
    \caption{5\,s}
	\label{fig:LED-interference:3}       % Give a unique label
	\end{subfigure}	
	\begin{subfigure}[c]{0.23\textwidth}
	\centering
	\includegraphics[width=\linewidth]{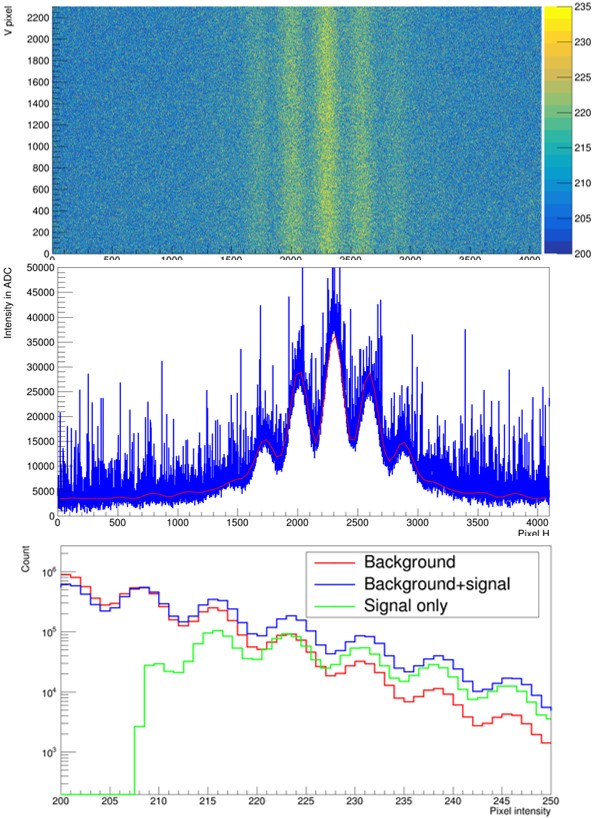}
    \caption{60\,s}
	\label{fig:LED-interference:4}       % Give a unique label
	\end{subfigure}	
    \caption{Single photon interference fringe measured by the camera with different exposure time: top row, the 2-D fringes; middle row, the 1-D intensity of fringe; bottom row, the 1-D intensity of pixels.}
	\label{fig:LED-interference}       % Give a unique label
\end{figure}

After the replacement of the PMT by the camera, a survey on exposure time is further performed from 172\,\textmu s to 60\,s as shown in Fig.\,\ref{fig:LED-interference}. The LED flashing frequency here is set to 2\,MHz, where there is still enough interval time between pulses to ensure only one photon in maximum 100\,ns duration traveling to the camera after the double-slit d2 inside the tube of the interferometer (length 60\,cm). The interference fringe is getting more clear when the exposure time increasing from 0.1\,s in Fig.\ref{fig:LED-interference:1} to 60\,s in Fig.\ref{fig:LED-interference:4}. The middle row of Fig.\ref{fig:LED-interference} is showing the light intensity projected to horizontal axis of each of the interference fringe, where the intensity of the interference fringe is gradually strengthening when exposure time increasing. According to the 1-D intensity plot of all the pixels of the camera (camera noise subtracted) as shown on the bottom row of Fig.\ref{fig:LED-interference}, it is clear that most of the pixels do not receive any photons (baseline at around 200\,ADC), few of the pixels receive only one photon (around 208\,ADC), and the maximum is around 2\, photons (around 216\,ADC) with 0.1\,s exposure time. Please note that, the photons can not arrive the camera at same time or single pulse duration according to the setting even single pixel can receive more than one photons. Both of the pixel number with more than 1\,photons and the maximum intensity of single pixel are increasing when the exposure time increasing as expected.

\iffalse
\begin{figure}[!htb]
	\centering
	\begin{subfigure}[c]{0.45\textwidth}
	\centering
	 \includegraphics[width=\linewidth]{figures/20220416-2100V-2MHz-100ns-interference-intensity-pulse.jpg}
	\caption{Intensity of each pulse}
	\label{fig:interference:intensity:pulse}       % Give a unique label
	\end{subfigure}	
	\begin{subfigure}[c]{0.45\textwidth}
	\centering
	 \includegraphics[width=\linewidth]{figures/20220416-2100V-2MHz-100ns-interference-intensity-pixel.jpg}
	\caption{Intensity of each pixel}
	\label{fig:interference:intensity:pixel}       % Give a unique label
	\end{subfigure}	
    \caption{Light intensity measured by Camera during the interference fringes testing with single photon. (a) the light intensity normalized to each pulse with or not considering the camera noise; (b)the light intensity normalized to each pixel with or not considering the camera noise.}
	\label{fig:interference:intensity}       % Give a unique label
\end{figure}
\fi

\begin{figure}[!htb]
	\centering
	\begin{subfigure}[c]{0.6\textwidth}
	\centering
	\includegraphics[width=\linewidth]{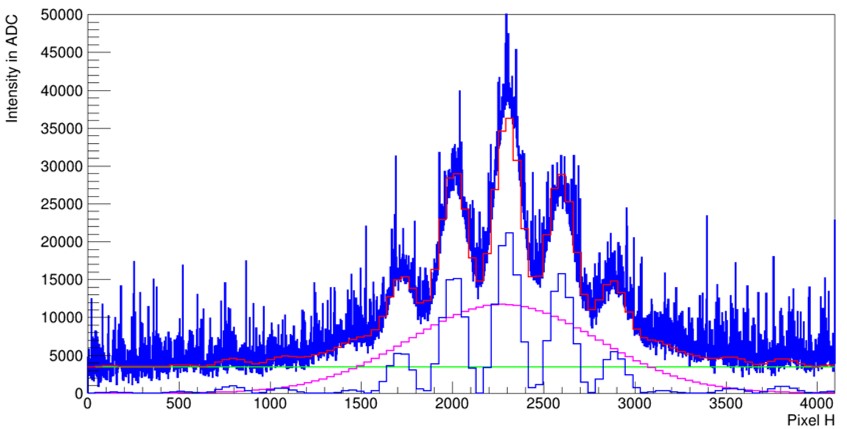}
	\caption{1-D plot fitting of the interference 60\,s}
	\label{fig:interference:intensity:1D}       % Give a unique label
	\end{subfigure}	
	\begin{subfigure}[c]{0.45\textwidth}
	\centering
	\includegraphics[width=\linewidth]{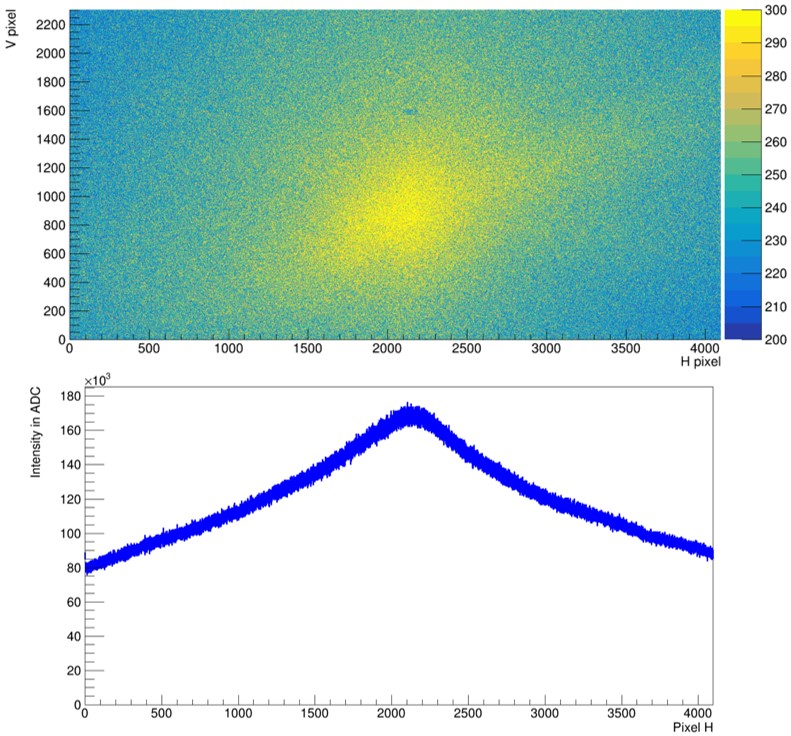}
    \caption{Bare LED without slits}
	\label{fig:interference:image:d2}       % Give a unique label
	\end{subfigure}
	\begin{subfigure}[c]{0.45\textwidth}
	\centering
	 \includegraphics[width=\linewidth]{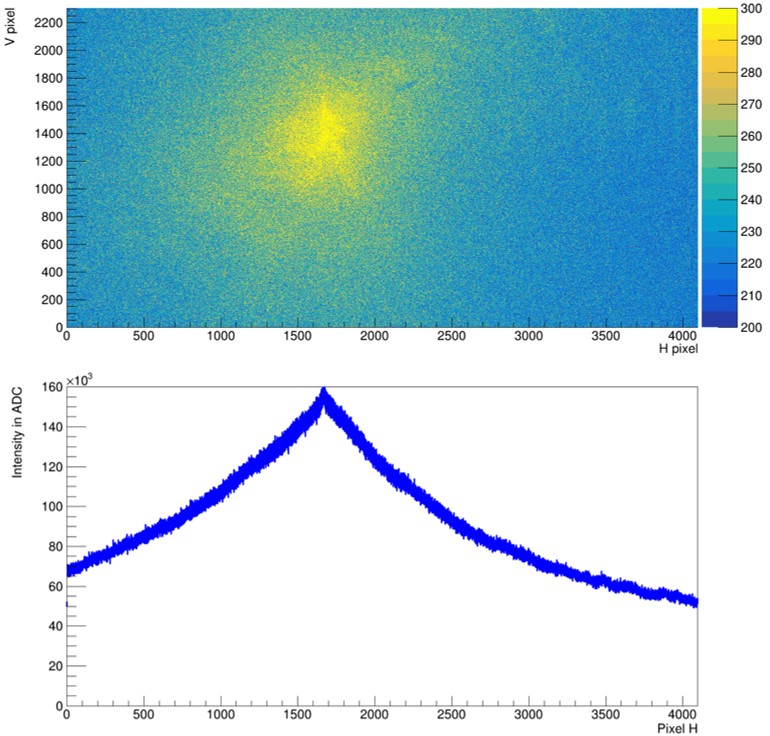}
    \caption{Only with slit d1}
	\label{fig:interference:image:d1}       % Give a unique label
	\end{subfigure}	
    \caption{(a) Fitting of the 1-D projection of the interference fringe with single photon and exposure time 60\,s; (b) the 2-D image (top) and its 1-D projection along horizontal direction (bottom) of the bare LED without any slits with 1\,s exposure time; (c) the 2-D image (top) and its 1-D projection along horizontal direction (bottom) only with the slit d1 with 5\,s exposure time.}
	\label{fig:interference:pattern}       % Give a unique label
\end{figure}

The projected intensity of the interference fringe with 60\,s exposure time is fitted following the intensity model from \cite{Feynman-1963} as shown in Fig.\,\ref{fig:interference:intensity:1D}, where a constant plateau and an extra Gaussian part are considered relative to the maximum line except the double-slit interference intensity model. The intensity of the constant plateau is related to the exposure time, which should be source from the subtraction of camera noise and the non-ideal interference system. The contribution of the Gaussian part should be related to the non-ideal interference system (such as the width difference of the slits of d2) and also observed in other double-slit experiments \cite{double-slit-PhysRevA.95.042129,double-slit-review-2002}. The measured wavelength of the LED is 477$\pm$2\,nm from the derived distance 0.213\,mm between the slits of d2 and the distance between the interference fringe 292\,pixels.
Fig.\,\ref{fig:interference:image:d2} shows the image of only with the LED by removing all the slits of d1 and d2 with a little lower light intensity and 1\,s exposure time, where we can identify a clear asymmetry along horizontal or vertical direction on the 2-D image, and the projecting 1-D intensity along horizontal direction.
Fig.\,\ref{fig:interference:image:d1} shows the image only with slit d1 with 5\,s exposure time, where the asymmetry distribution but different to the bare LED itself still can be identified on the 2-D image and the projecting 1-D intensity along horizontal direction. The asymmetry distribution could contribute to the constant and Gaussian part of the interference fringe after the double-slit d2.

\begin{figure}[!htb]
	\centering
	%\begin{subfigure}[c]{0.75\textwidth}
%	\centering
	 \includegraphics[width=0.6\linewidth]{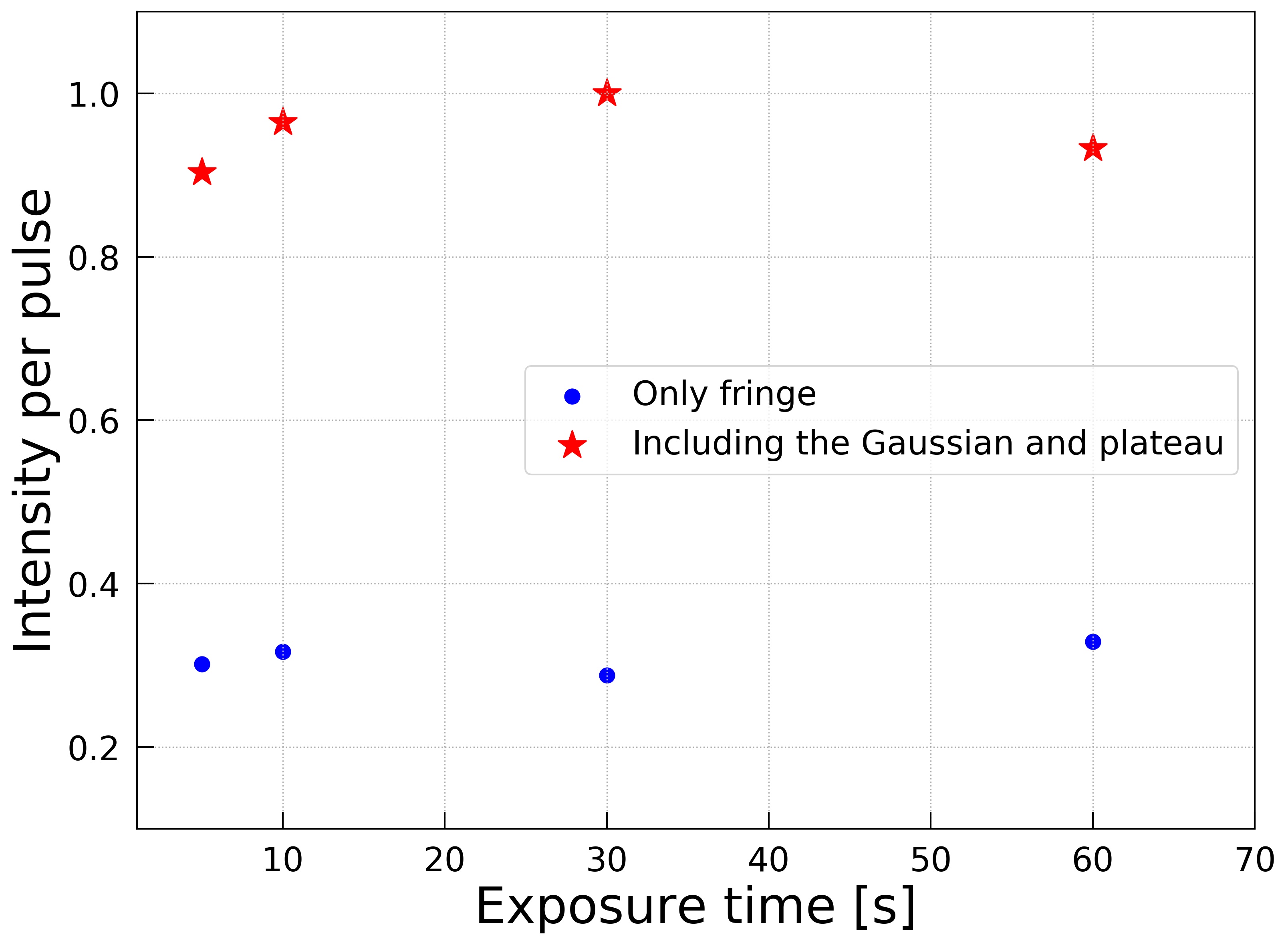}
%	\caption{Intensity of each pulse}
%	\label{fig:interference:intensity:pulse}       % Give a unique label
	%\end{subfigure}	
    \caption{Light intensity measured by the camera during the interference measurements with single photon considering or not the Gaussian and the plateau part, which is normalized to each LED pulse.}
	\label{fig:interference:intensity}       % Give a unique label
\end{figure}

As shown in Fig.\,\ref{fig:interference:intensity}, the light intensity on average measured by the camera is around 0.04\,p.e. per pulse only considering the fringe (0.12\,p.e. per pulse considering all the parts including the Gaussian and Plateau except camera noise), or 80,000\,p.e./s only considering the fringe (around 240,000\,p.e./s considering all the others except camera noise), respectively. It is basically consistent with the prediction by the PMT, and their difference is mainly from the QE and acceptance area, which still needs detailed validation.

\section{Imaging of crystal}
\label{sec:1:crystal}

%\subsection{Background}
%\label{sec:2:crystal:bkg}

A 7.5$\times$7.5$\times$15\,cm$^3$ CsI(Tl) crystal is put in front of the camera with a distance around 10\,cm (around 15\,cm to the PMT) as shown in Fig.\ref{fig:darkbox-crystal}, where the camera is coupled to another lens with much short focus length (1/2", C, 6-$\infty$\,mm, f/1.4). The 3-inch PMTs are used to monitor the signal intensity of the crystal, and the coincidence of the two PMTs is used as the trigger of the DT5751, where the threshold of each PMT is set to around 1\,p.e.

An $^{241}$Am alpha source (10\,kBq, side/surface source, used for smoke alarm with a diameter of around 4\,mm) is put on the top surface along horizontal direction of the crystal directly, and its rate is around 150\,count per second measured by a Geiger counter near to the source surface. Here it is considered the alpha particle only (energy around 5.5\,MeV), which is still the main contributor to energy deposition than the X-ray (59\,keV) even considering the quenching factor\cite{CsI-quenchingfactor-2002-ARK2002460,CsINa-neutron-2016-GUO201638,CsI-quenching-2015-LEE2015133}.

\begin{figure*}[!hbt]
    \centering
    \begin{subfigure}[c]{0.32\textwidth}
	\includegraphics[width=\linewidth]{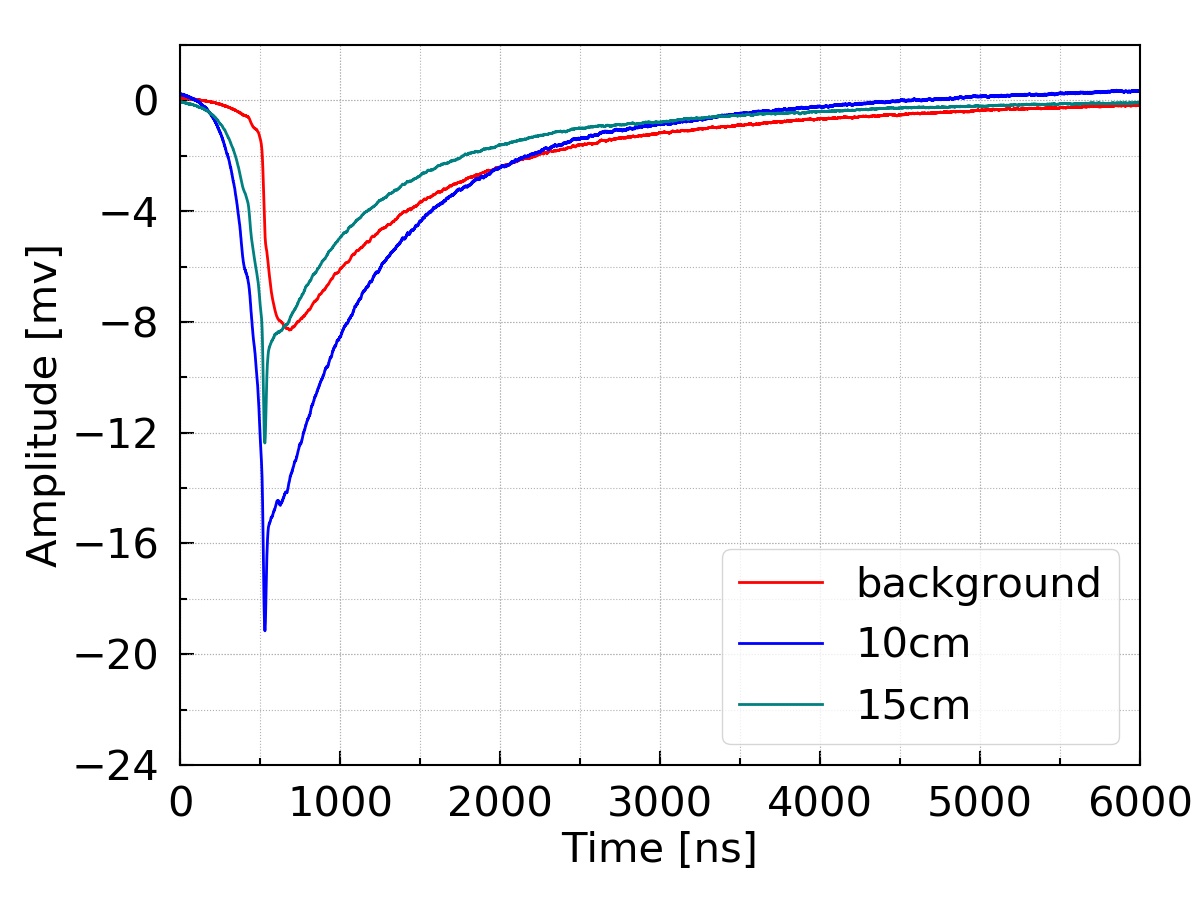}
    \caption{Average waveform}
	\label{fig:crystal:bkg:PMT:wave}       % Give a unique label
	\end{subfigure}	
	\begin{subfigure}[c]{0.32\textwidth}
	\includegraphics[width=\linewidth]{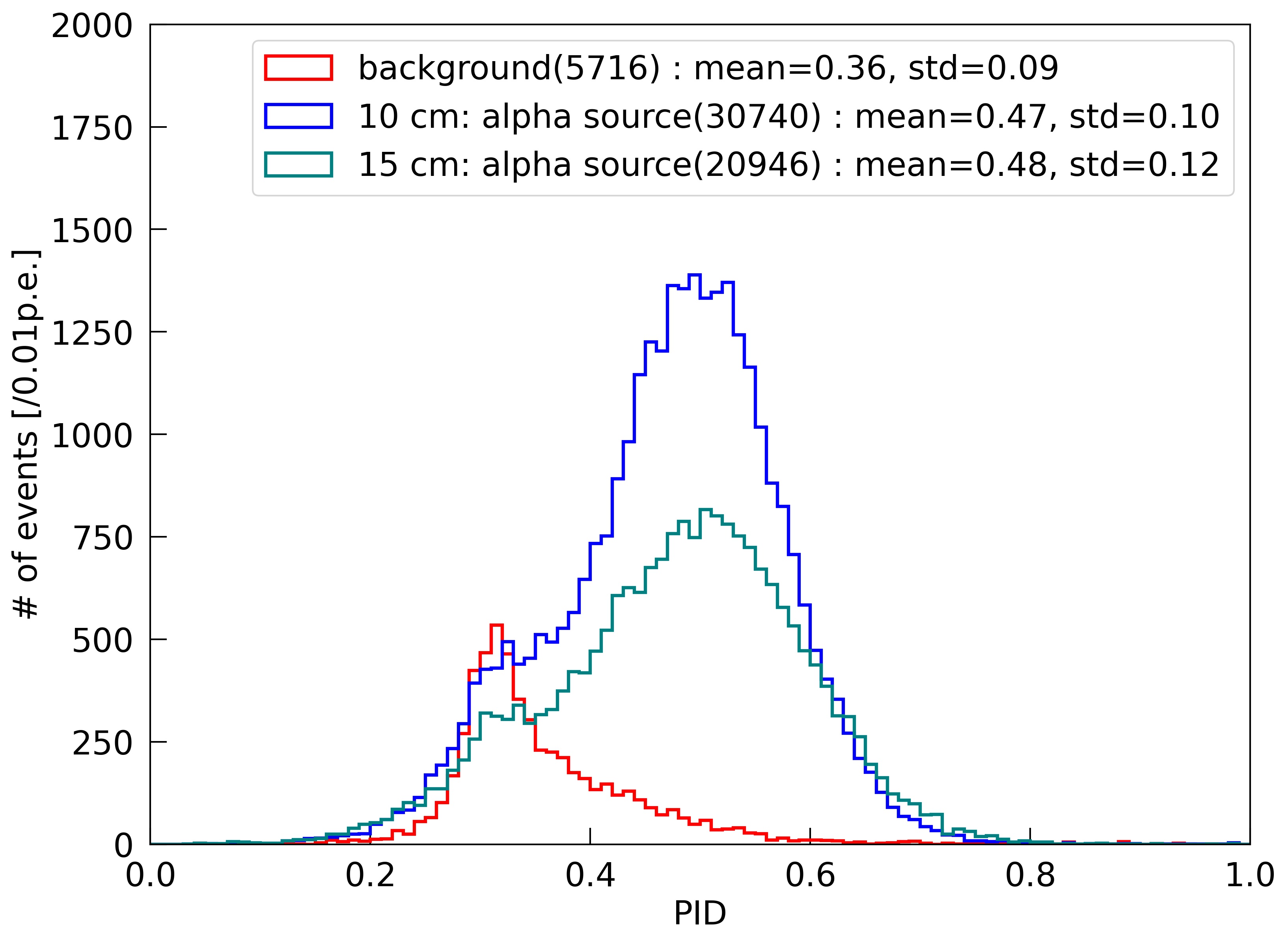}
    \caption{PID of crystal}
	\label{fig:crystal:bkg:PMT:PID}       % Give a unique label
	\end{subfigure}	
	\begin{subfigure}[c]{0.32\textwidth}
	\includegraphics[width=\linewidth]{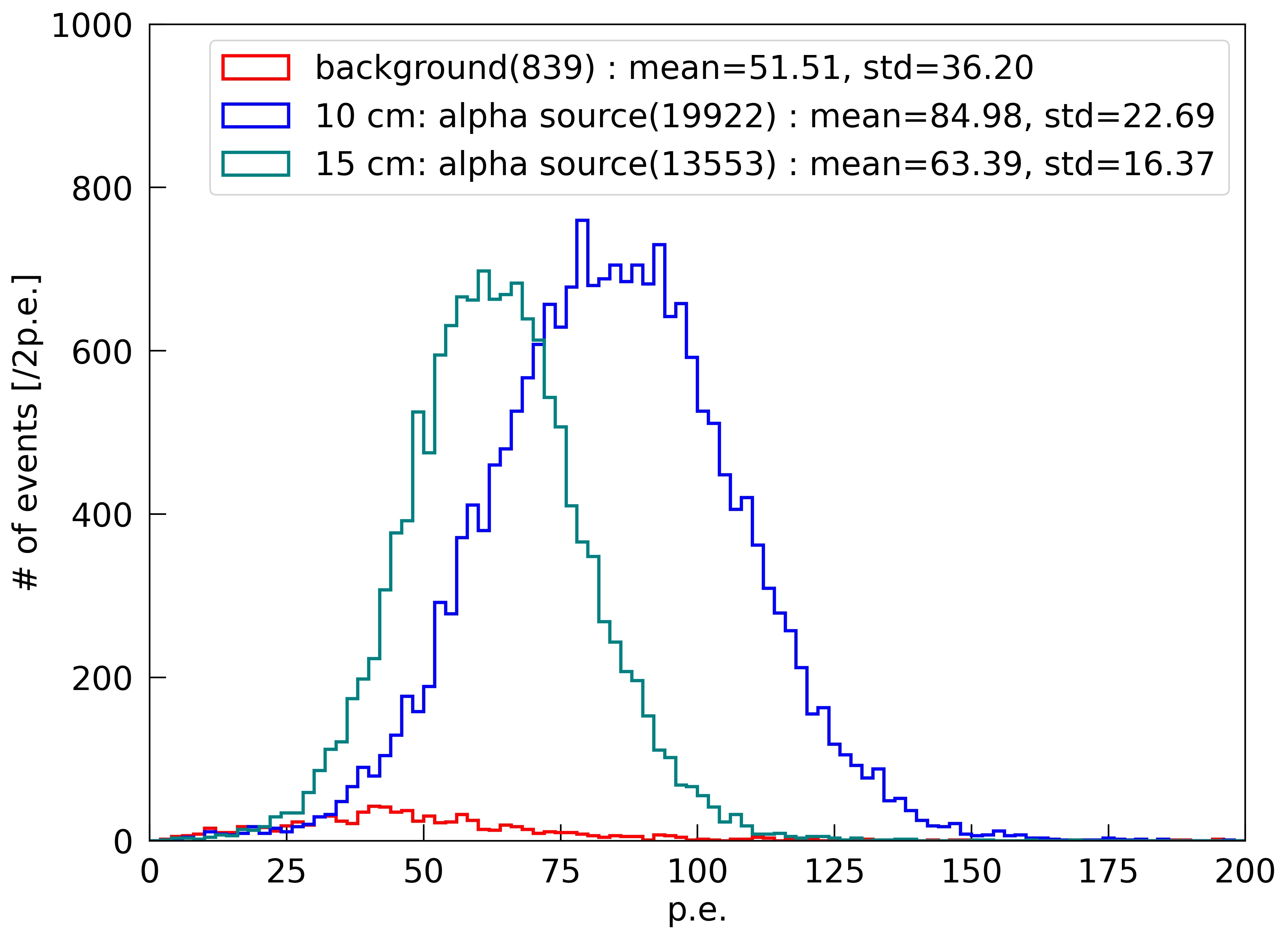}
    \caption{Charge intensity}
	\label{fig:crystal:bkg:PMT:charge}       % Give a unique label
	\end{subfigure}	
    \caption{The monitoring results by PMT of the crystal and alpha source.}
	\label{fig:crystal:bkg:PMT}       % Give a unique label
\end{figure*}

The monitored rate is around 10$\pm$1\,count per second only with the crystal by the data acquisition system, and around 130$\pm$10\,count per second with the source with the normalization according to the selected muon rate.
Fig.\ref{fig:crystal:bkg:PMT:wave} shows the averaged waveform from one of the monitoring PMTs, which demonstrates the slow fluorescent decay time in \textmu s of the CsI(Tl) crystal and the feature for different particles\cite{CsI-decaytime-MASUDA1992135,CsI-decaytime-SKULSKI2001759}. The PID of CsI(Tl) \cite{PID-CsI-BENRACHI1989137,PID-CsI-TERANISHI2021164967} is calculated for all the events with a fast signal window of 500\,ns as shown in Fig.\ref{fig:crystal:bkg:PMT:PID}, where the alpha events (selected by PID\,>\,0.45) can be identified clearly from the background, and the identified rate of muon and alpha is around 3.0$\pm$0.1\,Hz (by PID and total charge) and 90$\pm$10\,Hz at 10\,cm distance, respectively. The PMT measured charge of selected alpha events is further plotted as shown in Fig.\ref{fig:crystal:bkg:PMT:charge}, which considered the integration window of 6\,\textmu s. The typical intensity of single alpha event viewed by the PMT is around 80$\pm$2\,p.e. at 10\,cm distance to the PMT. It is expected to around 8$\pm$1\,p.e. per alpha event viewed by the camera, which is around 1/10 to that viewed by the PMT considering the solid angle of the lens, the QE, and lens transparency.

\begin{figure*}[!hbt]
    \centering
    \begin{subfigure}[c]{0.4\textwidth}
	\includegraphics[width=\linewidth]{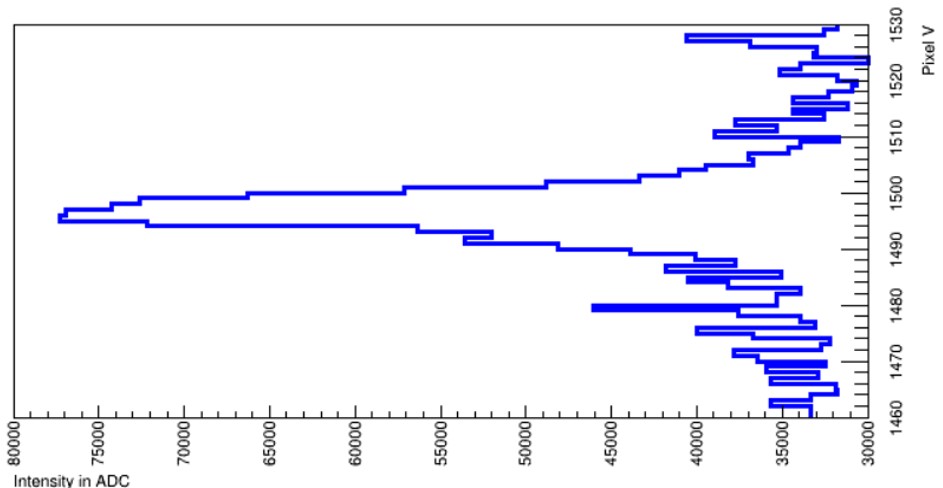}
    \caption{1-D intensity on vertical}
	\label{fig:crystal:bkg:2D:V}       % Give a unique label
	\end{subfigure}	
    \begin{subfigure}[c]{0.45\textwidth}
	\includegraphics[width=\linewidth]{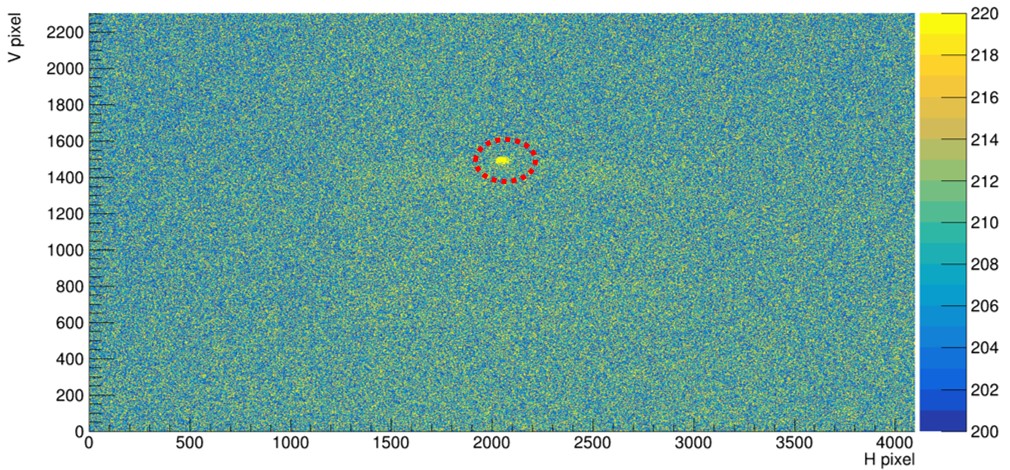}
    \caption{2-D image of crystal w/ alpha}
	\label{fig:crystal:bkg:2D}       % Give a unique label
	\end{subfigure}	
    \begin{subfigure}[c]{0.4\textwidth}
    \includegraphics[width=\linewidth]{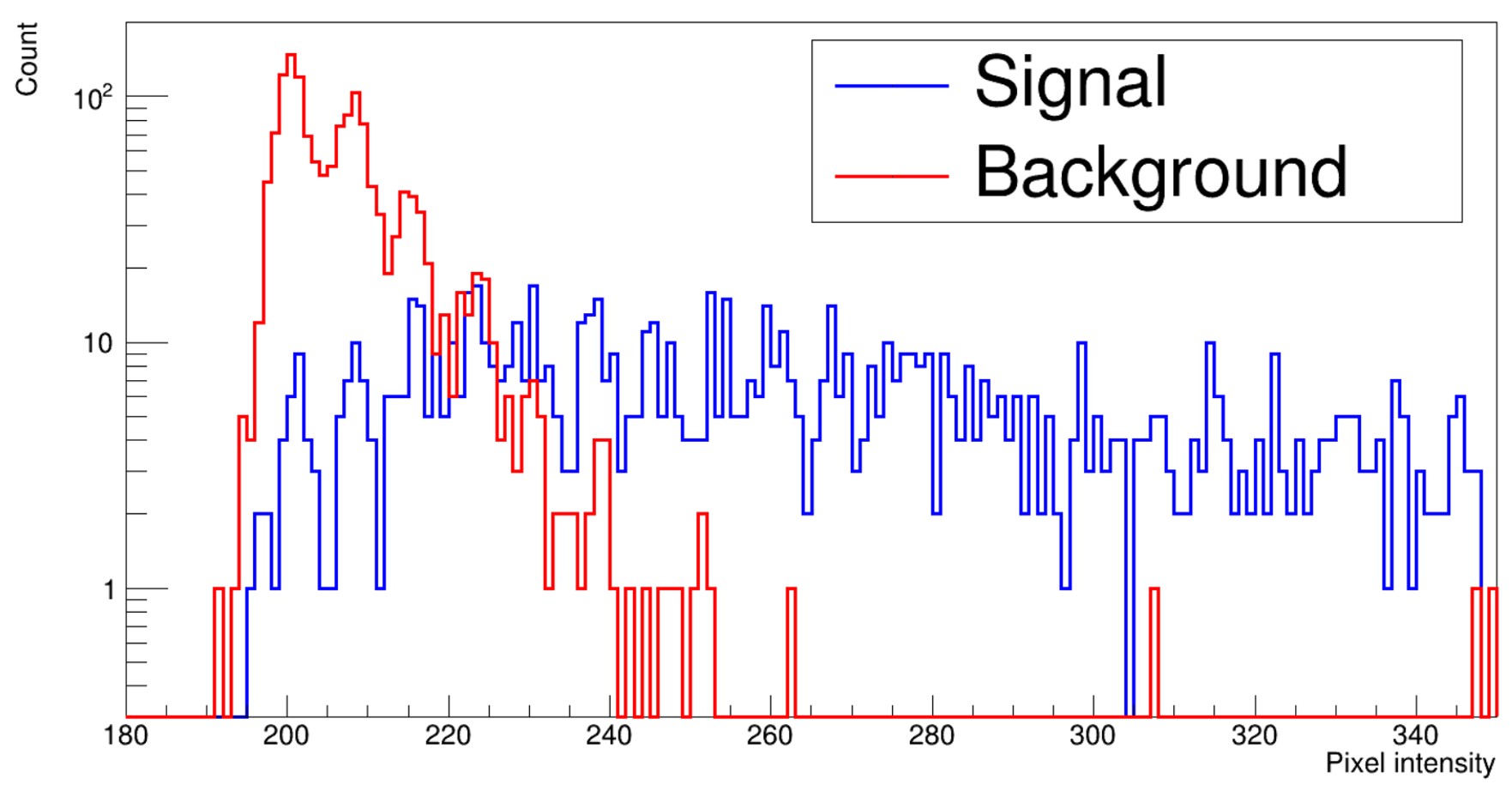}
    \caption{Pixel intensity of source region}
    \label{fig:crystal:bkg:1D}
    \end{subfigure}	
    \begin{subfigure}[c]{0.42\textwidth}
    \includegraphics[width=\linewidth]{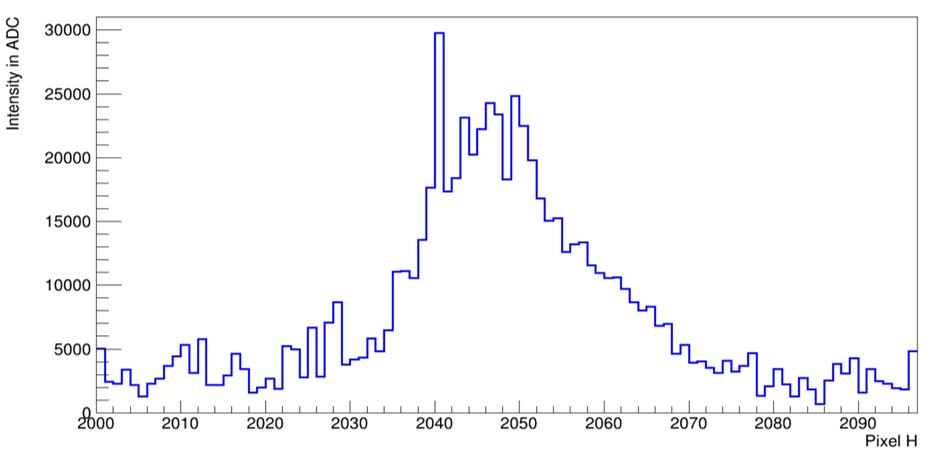}
    \caption{1-D intensity on horizontal}
    \label{fig:crystal:bkg:2D:H}
    \end{subfigure}
    \caption{Crystal and alpha source with 60\,s exposure time.}
	\label{fig:crystal:bkg}       % Give a unique label
\end{figure*}

A taken image with exposure time 60\,s is shown in Fig.\,\ref{fig:crystal:bkg:2D}, where the location of the alpha source can be identified clearly as shown in the dashed red line circle. According to the zoom in of the projected plot along horizontal as shown in Fig.\ref{fig:crystal:bkg:2D:H}, the source dimension along horizontal direction is around 45 pixels, which means 3.4\,mm considering the imaging magnification and basically consistent with the source dimension. According to the zoom in of the projected curve along vertical as shown in Fig.\ref{fig:crystal:bkg:2D:V}, the energy deposit depth in vertical direction is around 15 pixels, which means around 1.1\,mm considering the imaging magnification.

A radius of 22 pixels of the alpha source region (center at around pixel (2047,1497)) is used finally to calculate the signal intensity in the following discussion. At the same time, the shape of the crystal can be identified dimly against the discrete noise because of the reflection of the crystal surface.
The measured intensity of each pixel in the region of alpha source is plotted in Fig.\ref{fig:crystal:bkg:1D}, including a curve of the source region, and a curve of background from an shifted equal area far from the source. It is clearly seen that the typical intensity of each pixel of the source region is around 260\,ADC (baseline 200) or 8\,p.e., and the integrated intensity of the source region on this image is around 50,000\,p.e., which will be used as the measured signal strength.

\begin{figure*}[!hbt]
    \centering
    \includegraphics[scale=0.35]{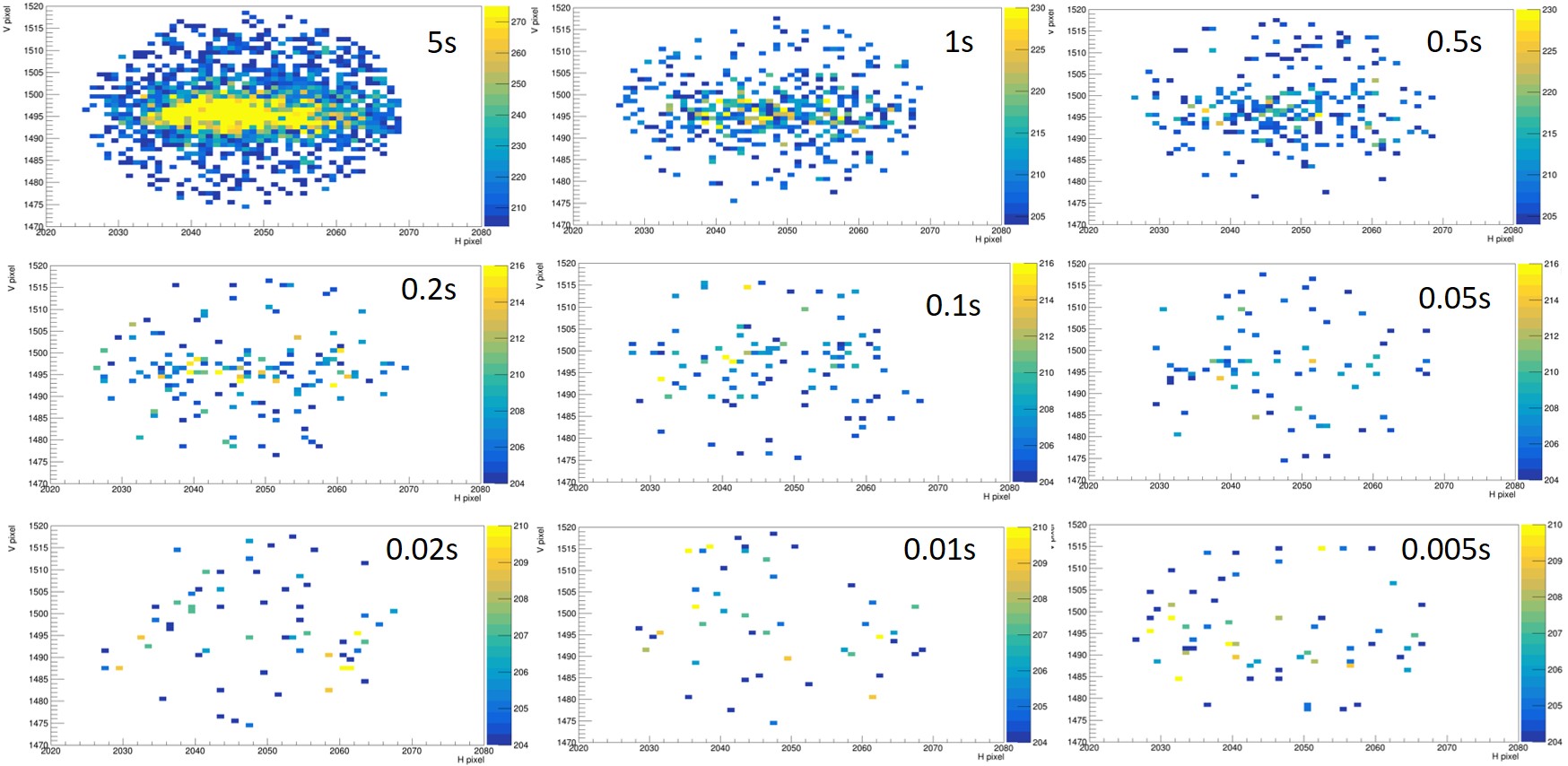}
    \caption{Images of crystal and alpha source with different exposure time, where the value of each pixel is in unit of ADC with 200\,ADC baseline.}
    \label{fig:crystal:alpha:multi}
\end{figure*}

%\subsection{With gamma source}
%\label{sec:2:crystal:gamma}

\begin{figure*}[!hbt]
    \centering
    \includegraphics[width=0.6\linewidth]{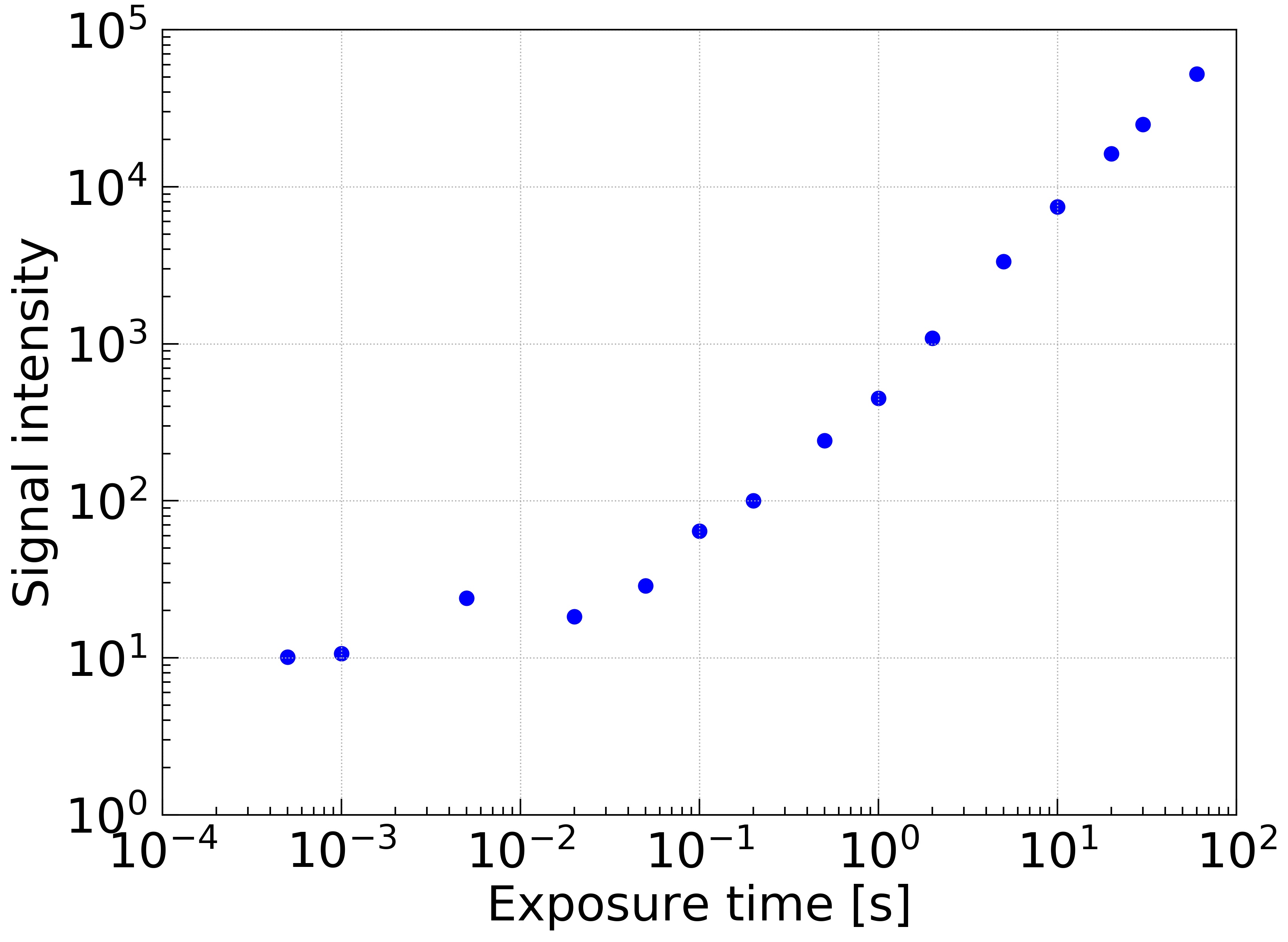}
    \caption{Signal intensity vs. exposure time.}
    \label{fig:crystal:alpha:intensity}
\end{figure*}

The alpha source region is further measured with different exposure time to derive the signal strength as shown in Fig.\ref{fig:crystal:alpha:multi}, where the destination region is zoomed in for detailed checking.
The significance of the source is becoming indistinct when shortening the exposure time from 5\,s to 0.01\,s, where it is not able to tell from the noise as with 0.01\,s or shorter exposure time (less possibility to catch an alpha event).
Fig.\ref{fig:crystal:alpha:intensity} shows the the measured signal intensity (with background subtraction) versus the exposure time, which gives a slope factor around 860\,p.e./s or 6700\,ADC/s, which means around 9.5\,p.e./alpha event.
The calculated signal intensity reaches a plateau at 100\,ADC or 12\,p.e., dominated by the fluctuation of camera noise in the target region, when the exposure time is smaller than 0.01\,s. It will only provide a signal-to-noise ratio around 0.8 if only single alpha event showing up now. It needs to further improve the signal-to-noise ratio by increasing the signal strength (including enlarge the effective aperture) or reducing the noise level (and smaller the target area).

\section{Discussion}
\label{sec:1:dis}

According to the previous measurements with the single photon sensitive camera, it is approaching to have an event by event imaging for the scintillation detector under the photon-starved regime and uniform angular distribution of the light produced in the scintillation process. Following the testing results of CsI(Tl) crystal and alpha source, it is valuable to maximum the effective aperture from the used lens with f/1.4 to a larger one, or to measure tracks with smaller and predictable target area to reach a better signal-to-noise ratio. Considering the noise of the camera, further reducing is still favored, even it is already smaller than 0.3\,p.e. per pixel revolutionary, .

\begin{figure*}[!hbt]
    \centering
    \includegraphics[scale =0.4]{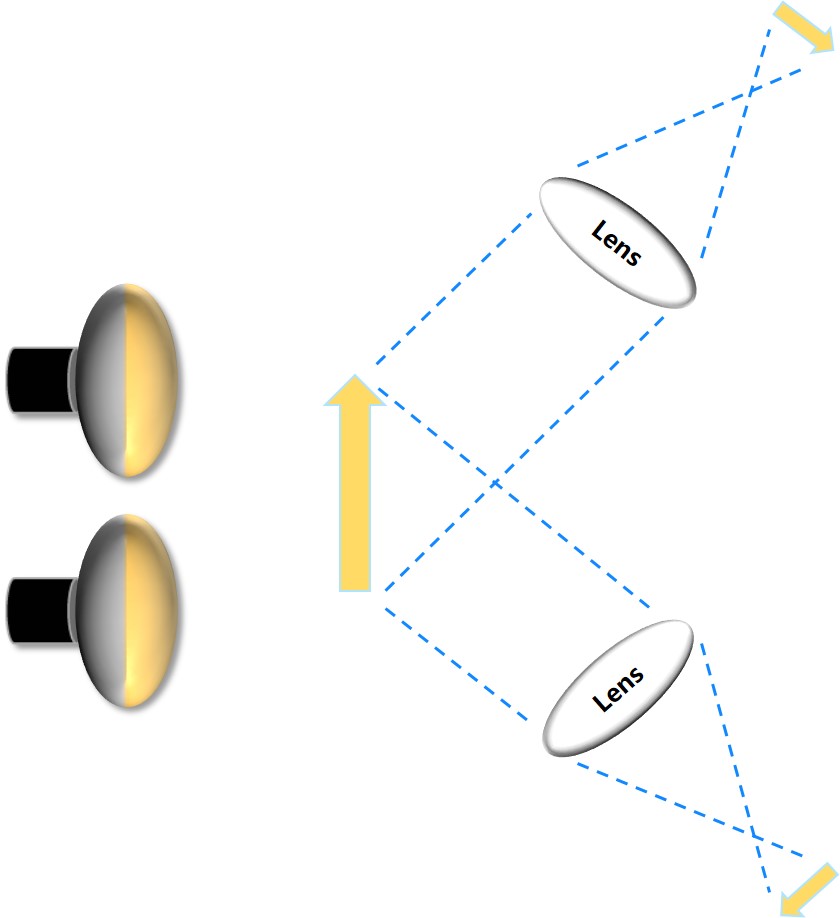}
    \caption{Schema of the camera spatial coincidence system with crystal.}
    \label{fig:spatial-coincidence}
\end{figure*}

At the same time, drawing lessons from the coincidence of PMTs in time, a spatial coincidence system with more than one single photon sensitive camera is also valuable and workable to suppress the noise, as proposed in Fig.\ref{fig:spatial-coincidence}. With the system, the PMTs mainly focus on a real time event measurement on timing, energy and crude spatial vertex. The cameras will help on precisely spatial vertex imaging, and the spatial coincidence among cameras could depress the random noise effectively.

\section{Summary}
\label{sec:1:summary}

The single photon sensitive and low noise camera provides another novel possibility on imaging for the photon-starved regime and uniform angular distribution of scintillation detectors. This study confirms the performance of the newly developed camera for photon counting with sub-photoelectron noise per pixel per second. With a combined system constituted by 3-inch PMTs and the camera, the double-slit Young's interference with single photon can be achieved, and single $^{241}Am$ alpha event with CsI(Tl) crystal is closely to be distinguished, where the camera with a 1/2", f/1.4 lens can collect around 1/10 of photons viewed by the 3-inch PMT. A spatial coincidence system is further proposed as an attractive option to depress the random noise of the camera and improve the signal-to-noise-ratio, where the lenses with larger effective aperture are also favored for further applications.

\section*{Acknowledgments}
\label{sec:1:acknow}

This work was supported by the National Natural Science Foundation of China No.\,11875282, the State Key Laboratory of Particle Detection and Electronics, SKLPDE-ZZ-202208. The authors would like to thank Mr. Dongsheng Shao for the valuable discussions and help on the optical lenses.

%\appendix
%\section{Some title}
%Please always give a title also for appendices.

%\acknowledgments

%This is the most common positions for acknowledgments. A macro is
%available to maintain the same layout and spelling of the heading.

%\paragraph{Note added.} This is also a good position for notes added
%after the paper has been written.

%\acknowledgments

%This is the most common positions for acknowledgments. A macro is
%available to maintain the same layout and spelling of the heading.

%\paragraph{Note added.} This is also a good position for notes added
%after the paper has been written.

\bibliographystyle{unsrtnat}
\bibliography{allcites}   % name your BibTeX data base

\end{document}